\documentclass[apjl]{emulateapj}
\usepackage{graphicx}

\slugcomment{submitted to The Astrophysical Journal (March 31, 2013)} 

\shorttitle{ALMA Observations of HH 46/47}
\shortauthors{Arce, et al.}

\begin{document}
%\doublespace

\newcommand{\kms}{km\ s$^{-1}$}
\newcommand{\cc}{cm$^{-3}$}
\newcommand{\co}{$^{12}$CO}
\newcommand{\vchan}{$v_{chan}$}
\newcommand{\vout}{$v_{out}$}
\newcommand{\venv}{$v_{env}$}

\title{ALMA Observations of the HH 46/47 Molecular Outflow }
%\author{Arce, Mardones, Corder, et al.}
\author{H\'ector G.~Arce}
\affil{Department of Astronomy,Yale University, P.O.~Box 208101, New Haven, CT 06520-8101, USA} 
%\email{hector.arce@yale.edu} 
\author{Diego Mardones}
\affil{Departamento de Astronom\'{i}a, Universidad de Chile, Casilla 36-D, Santiago, Chile}
%%\email{diego@das.uchile.cl}
\author{Stuartt A.~Corder}
\affil{Joint ALMA Observatory, Av. Alonso de C\'ordova 3107, Vitacura, Santiago, Chile}
%%\email{scorder@alma.cl}
\author{Guido Garay}
\affil{Departamento de Astronom\'{i}a, Universidad de Chile, Casilla 36-D, Santiago, Chile}
%%\email{guido@das.uchile.cl }
\author{Alberto Noriega-Crespo}
\affil{Infrared Processing and Analysis Center, California Institute of Technology, Pasadena, CA 91125, USA}
%%\email{alberto@ipac.caltech.edu}
\and
\author{Alejandro C. Raga}
\affil{Instituto de Ciencias Nucleares, UNAM, Ap.~70-543, 04510 D.F., M\'exico}
%%\email{raga@nucleares.unam.mx}
%\and
%\author{Sylvie Cabrit}
%\affil{LERMA, UMR 8112 du CNRS, Observatoire de Paris, Ecole Normale Sup\'erieure, Universit\'e Pierre et Marie Curie, Universit\'e de Cergy-Pontoise, 61 Av. de lÕObservatoire, 75014 Paris, France}
%%\email{sylvie.cabrit@obspm.fr}

%\shorttitle{HH46/47 Molecular Outflow}
%\shortauthors{Arce et al.}

\begin{abstract}
The morphology, kinematics and entrainment mechanism 
of the HH 46/47 molecular outflow were studied using  new ALMA Cycle 0 observations. 
Results show that the blue and red lobes are strikingly different. 
We argue that these differences are partly due to contrasting ambient densities 
that result in different wind components 
having a distinct effect on the entrained gas in each lobe.
A 29-point mosaic, covering the two lobes  at an angular resolution of about $3\arcsec$, detected outflow emission at much higher velocities than previous observations, resulting in significantly higher estimates of the outflow momentum and kinetic energy than previous studies of this source, using the CO(1--0) line. The morphology and the kinematics of the gas in the blue lobe are consistent with models of outflow entrainment by a wide-angle wind, and a simple model describes the observed structures in the position-velocity diagram and the velocity-integrated intensity maps. The red lobe exhibits a more complex structure, and there is evidence that this lobe is entrained by a wide-angle wind and a collimated episodic wind. Three major clumps along the outflow axis show velocity distribution consistent with prompt entrainment by different bow shocks formed by periodic mass ejection episodes which take place every few hundred years. 
Position-velocity cuts perpendicular to the outflow cavity show  gradients where the velocity increases towards the outflow axis, inconsistent with outflow rotation.
Additionally, we find  evidence for the existence of a small outflow driven by a binary companion. 

\end{abstract}

\keywords{ISM: jets and outflows --- stars: formation --- ISM: Herbig-Haro objects --- ISM: individual (HH 46, HH 47)}

\section{Introduction}
\label{intro}
As stars form inside molecular clouds, they eject mass in energetic bipolar outflows. The resulting bipolar wind from a young stellar object (YSO) may reveal itself through Herbig-Haro (HH) objects observed in the optical, H$_2$ emission knots in the infrared (IR), and molecular (CO) outflows observed at millimeter (mm) wavelengths. HH objects delineate highly collimated jets and their (internal or leading) bow shocks. The H$_2$ IR emission also arises from recently shocked gas and in many cases it traces the bow-shock wings that extend toward the driving source. CO outflows map the ambient gas that has been swept-up well after it has been entrained by the protostellar wind and has cooled. Hence, these different manifestations provide complementary views: while HH objects and H$_2$ emission provide a ``snapshot'' of the current shock interaction, the CO outflow trace the protostar's mass loss history \citep[e.g.,][]{Richer+00}.

Protostellar winds inject energy and momentum into the surroundings, thereby perturbing the star-formation environment. Outflows may be responsible for the clearing of material from the core \citep{AS06}, a process that could result in the termination of the infall phase \citep[e.g.,][]{VL98}, affect the star formation efficiency in the cloud 
\citep[e.g.,][]{MM00, NL07, MH13}, and determine the mass of stars \citep{Myers08}. In addition, outflows can affect the kinematics, density and chemistry of a substantial volume of their parent clouds, and thus can be important to the turbulent dynamics and energetics of their host cores. How exactly protostellar winds entrain and disperse the surrounding gas and feed their parent cloudsÕ turbulence remains a mystery. High angular (and velocity) resolution, multi-wavelength, observations of outflows are needed to probe all different outflow manifestations in order to compare them with hydrodynamic numerical simulations and shock models to obtain a complete picture of protostellar winds, the entrainment process and their impact on their surroundings.

Here we present  Atacama Large Millimeter/sub-millimeter Array (ALMA) observations of the HH 46/47 molecular outflow. The high resolution data were obtained with the aim of studying the entrainment process, the underlying protostellar wind and the properties of the molecular outflow. The observed outflow is driven by HH 47 IRS (a.k.a., HH 46 IRS 1, IRAS~08242-5050), a low-mass Class I YSO  with a total luminosity of about 12 L$_{\sun}$ in the Bok globule ESO 216-6A, located on the outskirts of the Gum Nebula at a distance of 450 pc  \citep{Schwartz77, Reipurth+00, Noriega-Crespo+04}. The source lies very close to the edge of the Bok globule, which explains why the blue (northern) lobe of the HH 46/47 bipolar flow is clearly seen at optical wavelengths outside the globule, 
while within $\sim 2\arcmin$ of the source the red lobe (which mostly lies inside the
globule) is best seen in IR images. 
 Hubble Space Telescope (HST) observations indicate that HH 47 IRS is actually a binary system, where the two components of the system are observed to be separated by only 0.\arcsec26 or about 120 AU \citep{Reipurth+00}.  
   Wide-field, narrow-band H$\alpha$ and [S II] optical images of the region by \citet{Stanke+99}
reveal two groups of HH objects at  distances, from the position of HH 47 IRS,  
of about 1.3 pc to the northeast (in the blue lobe)
and southwest (in the red lobe), outside the parent globule. These results showed that 
the well-known HH 46/47 flow 
(extending 0.57 pc from HH 47D in the northeast to HH 47C to the southwest) is, in fact,  
the innermost part of 
a giant HH flow that extends 2.6~pc on the plane of the sky.

Many of the properties of the HH 46/47 flow have been determined through extensive optical and IR  observations. 
The combination of optical spectral data and  proper motion studies of the HH knots (using ground-based telescopes) allowed
 an estimate of the flow's inclination to the plane of the sky ($\sim 30\arcdeg$) and an average jet velocity  of 300~\kms \/ \citep{Reipurth89,RH91, EM94, Micono+98}. More recently, proper motion studies (using HST) of the blue lobe optical HH knots, in combination with the results from spectroscopic Fabry-Per\'ot observations \citep{Morse+94}, allowed \citet{Hartigan+05} to estimate an average orientation angle, with respect to the plane of the sky, of about $37\arcdeg$. 
  Infrared images show how shocked H$_2$ emission in the southwestern (redshifted) lobe traces the walls of a 36\arcsec \/  (0.08 pc) wide cavity that extends $\sim 2\arcmin$ (0.26 pc) to the position of HH 47C 
  \citep{Eisloffel+94, Noriega-Crespo+04}.

 The HH 46/47 molecular outflow has only been fully mapped at low resolution (with beams larger than 24\arcsec) by several authors, using the two lowest rotational transitions of CO \citep{CM91,Olberg+92,vanKempen+09}. APEX observations of higher CO transitions (up to 7-6) of a smaller area ($80\arcsec \times 80\arcsec$) surrounding the source show the molecular outflow gas reaches temperatures as high as 100 K \citep{vanKempen+09}. Moreover, recent Herschel PACS observations of a region within 30\arcsec \/ from the source reveal the existence of hot CO and H$_{2}$O (thought to be partly produced by non-dissociative C-shocks) and OH and [OI] emission from dissociative J-shocks where the protostellar wind interacts with the surrounding dense core \citep{vanKempen+10, Wampfler+10}.
 
Even though the extensive multi-wavelength studies of this source have helped understand the physics of protostellar jets, the lack of high-resolution molecular outflow observations (until now) has hampered our ability to  obtain a complete picture of the outflow phenomenon in HH 46/47, the entrainment process and the protostellar wind's impact on the cloud. Here we present the first interferometric, high-resolution 
(beam $\sim 3\arcsec$) 
observations of the HH 46/47 molecular (CO) outflow and compare the morphology and kinematics to existing (simple) models of outflow entrainment. In a future paper we will use  hydrodynamical simulations to model the outflow and aim to provide more stringent constraints on the launching and entrainment mechanisms in this flow.

 \begin{figure*}
\epsscale{0.9}
%\plotone{f1.eps}
\plotone{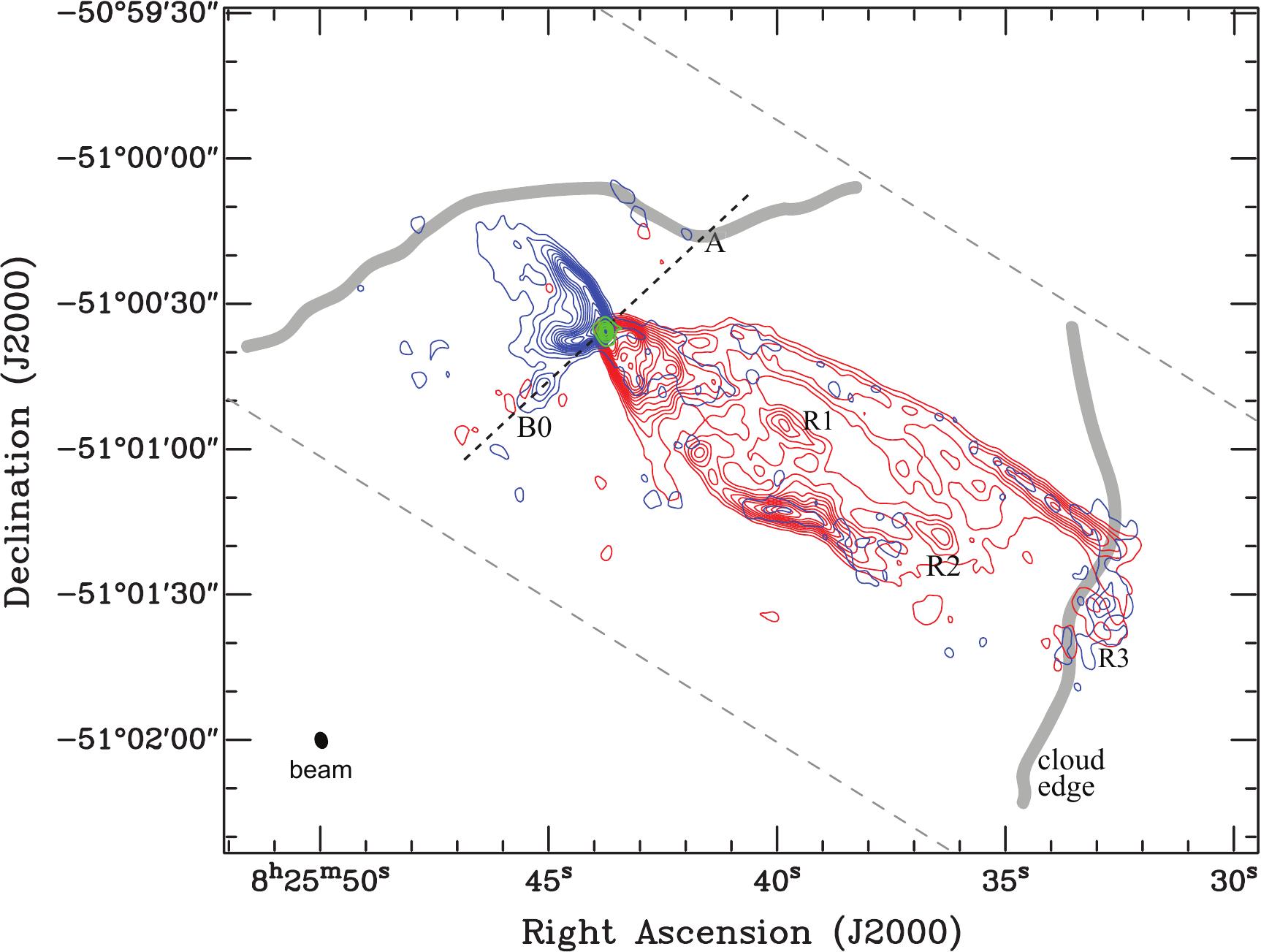}
%\vspace{-3in}
\caption{Integrated intensity map of the HH 46/47 CO(1--0) outflow. Blue (red) contours represent the 
 blueshifted (redshifted) lobe. The blue lobe is integrated over 
 $-6.8 < v_{out} < -1.3$~\kms. The lowest contour and subsequent contour steps are 0.2  and 0.25 Jy beam$^{-1}$~\kms, respectively. The red lobe is integrated over 
 $1.3 < v_{out} < 6.7$~\kms. The lowest contour and subsequent contour steps are 0.25  and 0.3 Jy beam$^{-1}$~\kms, respectively. Green contours show the 3 mm 
 continuum emission surrounding HH 47 IRS. The lowest contour and subsequent contour steps are 1.3  and 1 mJy beam$^{-1}$, respectively. The thick grey line delineates the edge of the globule as traced by our CO data close to the cloud velocity (see Figure~\ref{chanmapfig}). The synthesized beam of the CO map is 
 shown on the lower left corner of the figure. 
Grey diagonal parallel dashed lines show the edges of our map. The position of redshifted outflow clumps R1, R2 and R3 is shown, as well as that of  the blueshifted clump B0 and clump A.
The dark dashed line passing through B0, HH 47 IRS and A indicates the position of the $p-v$ cut shown in 
 Figure~\ref{pvbinary}.   
\label{outflowmap}}
\end{figure*}

\section{Observations}
\label{obs}
The  observations were carried out using ALMA from 28 December 2011 to 25 January 2012, during the Early Science Cycle 0 phase. 
The Band 3 data were obtained over eleven scheduling blocks, with 16 to 18 antennas in the (Cycle 0) 
compact configuration, consisting of projected baselines
in the range of 12 to 277 m.
The correlator was configured to observe three windows with the highest spectral resolution ($\delta\nu=30.5$ kHz),  with a bandwidth of 58.6 MHz and 
centered at  115.27 GHz, the frequency of the CO (1-0) transition, and at 
100.88 and 100.07 GHz. We only detected line emission in the 115.27 GHz window, which has a  velocity resolution %($\delta v$) 
of 0.08~\kms. No line was detected in the other two windows, 
and were therefore used to map the continuum emission. 
We mapped our sources using  a rectangular 29-point mosaic, with contiguous pointings separated by 24.5\arcsec, oriented at a position angle (P.A.) of
about $58\arcdeg$, designed to cover the length of the HH 46/47 outflow from HH 47A, in the northeast, to HH 46C, in the southwest, 
and the width of the red outflow lobe seen in IR images \citep[see, e.g.,][]{Noriega-Crespo+04}.  
The mosaic had three rows, the two outer rows consisted of 10 pointings and the central row had 9, resulting in
a map of about $1.2\arcmin   \times 4.2\arcmin$, centered on $08^h25^m41^s.5$, $-51\arcdeg00\arcmin47\arcsec$ \/ (J2000). 

The nearby quasar J0845-5458 was used for
phase and gain calibration. Flux calibration was carried out
using observations of Mars, and  J0538-440 and 3c279 were used as bandpass calibrators.
The visibility data were edited, calibrated and imaged (using the CLEAN algorithm) in CASA, using 
0.5\arcsec \/ cells and 1024 cells in each spacial dimension.
% We  used 0.5\arcsec \/ cells, 1024 cells in each spacial dimension and  XX weighting. 
 For the 
 spectral data we defined a different clean region for each channel, encircling the area with the brightest emission. 
The resulting synthesized beam for the $^{12}$CO(1--0)  data cube is $3.2\arcsec \times 2.4\arcsec$, with a PA$ = 16.5\arcdeg$, and 
the ($1 \sigma$) 
rms  is 25 mJy beam$^{-1}$ in a (spectroscopically smoothed) map  with a velocity resolution of 0.5~\kms.
The continuum was obtained by averaging the emission over all channels in the two windows close to 100 GHz with no detected line emission, resulting
in a bandwidth of 117.2 MHz. The synthesized beam and rms for the continuum map are $3.1\arcsec \times 2.2\arcsec$ (PA$ = 15\arcdeg$), and 0.33 mJy beam$^{-1}$

\section{Results}
\label{res}

\subsection{Continuum Source}
\label{contres}

We detected continuum thermal dust emission at 100 GHz near the position of the protostar, with a 
peak intensity of 8.1 mJy beam$^{-1}$.
A Gaussian fit puts the continuum peak at  $8^h25^m43.^{s}8, \/ $-51\arcdeg00\arcmin36\arcsec \/ (J2000.0),
which we adopt as the position of HH 47 IRS, and it is consistent with that given by  \citet{Reipurth+00}.
The continuum emission is barely resolved, and it is slightly extended toward the southwest 
(see Figure~\ref{outflowmap}). 
The total flux density is 14 mJy, and we follow \citet{Schnee+10}  and \citet{Dunham+12} to estimate the 
mass of the envelope using the dust continuum emission, assuming a dust temperature ($T_d$) of 30 K and gas-to-dust ratio of 100. 
We estimate the dust opacity ($\kappa$)
at 3~mm to be  0.9~cm$^2$~g$^{-1}$,
by extrapolating the value of $\kappa$ at 1.3~mm  
obtained by \citet{OH94} for dust with a thin ice mantle after $10^5$ yr of coagulation at a gas density of $10^6$~cm$^{-3}$, and assuming  an emissivity spectral index ($\beta$) of 1. 
Using the above assumptions,   
we obtain a total (gas and dust) mass associated with the continuum emission of 0.4 M$_{\sun}$. 
This estimate depends mostly on the assumed dust temperature, opacity and emissivity index. 
Single dish observations of the dust continuum of HH 47 by \citep{vanKempen+09} and observations of other Class 0 and Class I sources \citep[e.g.,][]{AS06} indicate that $T_d$ in the range of 20 to 40~K is a reasonable assumption. The dust opacity at 3~mm can vary depending on the adopted dust model and the
assumed $\beta$ (which may range from about 0.3 to 1.5 for Class 0 and Class I sources, e.g., 
Arce \& Sargent 2006). 
Reasonable variations of these quantities lead to possible variations in the mass estimate
by a factor of two to three.

The angular resolution of our continuum is not enough
 to resolve different peaks associated with the two different binary components of the system. The observed continuum emission most probably traces  the envelope that surrounds the binary system, and observations with an order of magnitude higher angular resolution are needed to investigate whether each component has its own separate circumstellar envelope.  

\subsection{Line Data and the Molecular Outflow}
\label{outflowres}
Integrated intensity maps of the CO(1--0) blueshifted and redshifted emission from the HH 46/47 molecular outflow are displayed in Figure~\ref{outflowmap}. 
The most striking aspect of this  outflow is the clear difference in size and morphology between the two different lobes; the red (southwestern) lobe extends 
approximately 2\arcmin \/ from the source, while the blue (northeastern) lobe only extends up to about 30\arcsec \/ from the powering YSO.
This asymmetry had been noted in earlier, lower angular resolution, observations of the CO outflow 
and it is generally assumed that 
it is due to the fact that the HH 47 IRS  is close to the edge of the parent globule \citep[i.e,][]{CM91,Olberg+92,vanKempen+09}. 
The  protostellar wind's blueshifted (northeastern) lobe  breaks-out of the cloud, 
where there is little molecular gas for it to entrain, while
the redshifted lobe dives into the globule and is able to entrain much more gas along its path. 

In addition to the drastic difference in length, each lobe exhibits dissimilar morphologies most likely caused by differing dominant entrainment processes. 
The blue lobe shows mainly a parabolic morphology, with a P.A. \/ of about $60\arcdeg$. The morphology and the velocity structure of the blue lobe suggests this molecular outflow is mostly being formed by the entrainment of cloud material by a wide angle protostellar wind (see~Section~\ref{bluelobe}). 
We also detect in the blue lobe a small protuberance southeast of the continuum source (labeled B0 in Figure~\ref{outflowmap}), which we argue is due to the outflow from a binary companion (see~Section~\ref{binary}). The red lobe, on the other hand, has a V-shape near the source, with an opening angle of about $65\arcdeg$. Starting at about 30\arcsec \/ from the source, the integrated intensity contours mostly trace the southern and northern outflow cavity walls, which extend up to  80\arcsec \/ and 2\arcmin, respectively, from the source. We suspect that these structures trace the limb-brightened walls of an approximately cylindrical shell 
produced by the outflow-cloud interaction. 
In addition, we detect three distinct clumps of gas  along the outflow axis
 which we label R1, R2 and R3 (see Figure~\ref{outflowmap}). In Section~\ref{redlobe}, we argue that these result from
approximately periodic  mass ejections driven by the central protostar. 
%We also detect faint blueshifted emission in three main 
%regions of the (predominantly) red lobe: near the outflow source (where the outflow shows a conical structure);  in the bright region of the of the southern outflow cavity wall (SW1?); and near R3.

\begin{figure}
\epsscale{1.0}
%\plotone{f2.eps}
\plotone{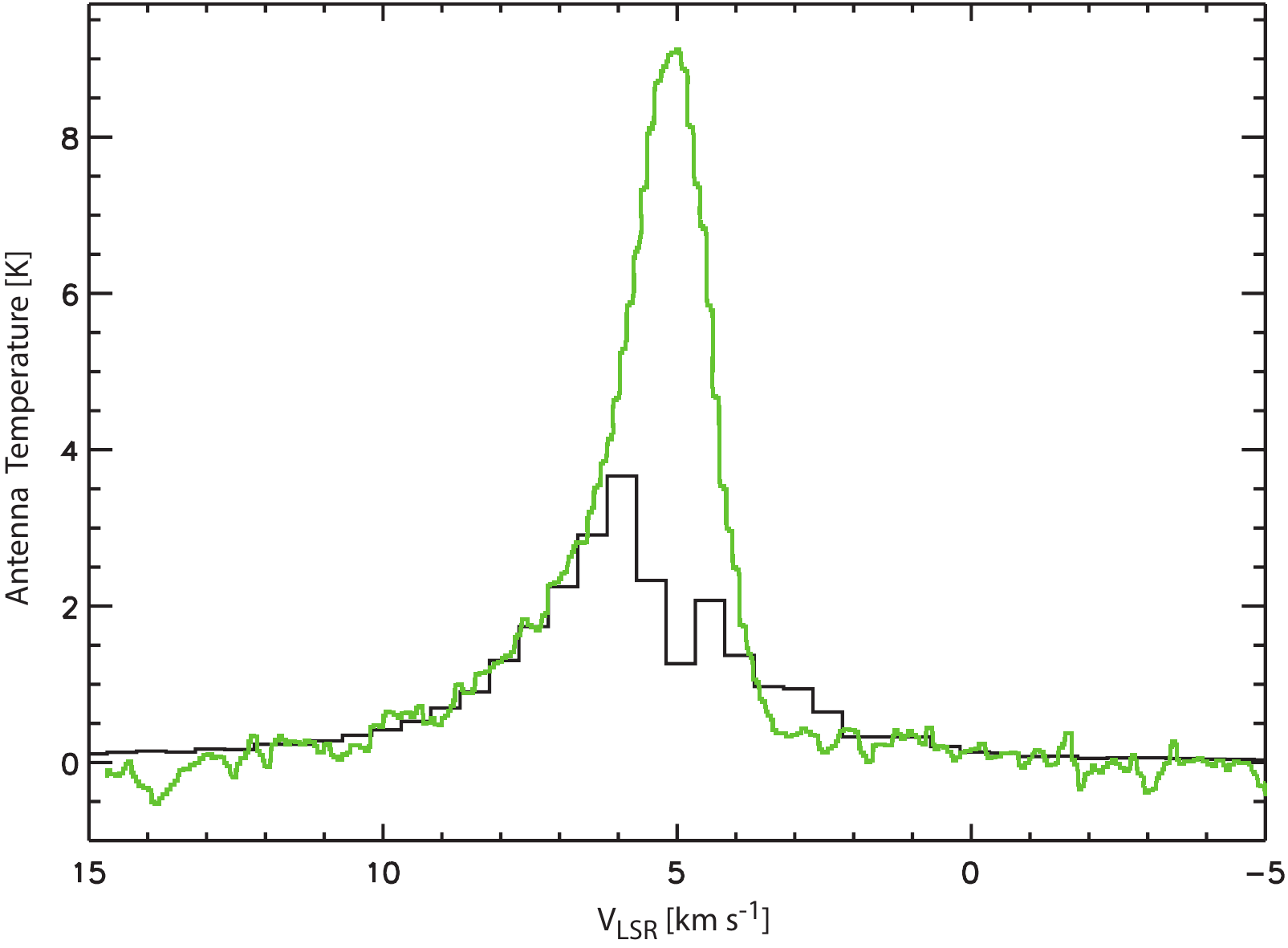}
\caption{Comparison between CO(1--0) spectrum from \citet{Olberg+92} and our ALMA observations. 
 The spectrum shown in green is the same one shown in  Figure 4 of \citet{Olberg+92}, which comes
 from  the brightest position in the redshifted lobe of their the map. 
The black histogram shows the spectrum at the same position from a map using our ALMA data that have been smoothed to the same beam size as that of the Olberg et al.~data (44\arcsec). The spectrum from the ALMA data is also smoothed to a velocity resolution of 0.5~\kms \/ for easier visual comparison with the Olberg et al.~spectrum.
\label{avespec}}
\end{figure}

%In Figure~\ref{avespec} compare our data to a spectrum that of CO(1--0) data from \citet{Olberg+92}, by plotting the spectra from the same position   
In Figure~\ref{avespec} we plot the spectrum 
at the position of  brightest integrated intensity in the CO(1--0) outflow map of \citet{Olberg+92}. We compare the spectrum shown in Figure 4 of \citet{Olberg+92} with a spectrum obtained from our CO map, smoothed to  the same resolution as the beam size of the Olberg et al.~observations (44\arcsec). From this comparison we see that the ALMA observations recover most (if not all) of the emission at 
LSR velocities greater than 6.5~\kms \/ and less than 3.5~\kms, and it is clear that we do not fully recover the emission at velocities close
to that of the cloud ($v_{LSR} \sim 5$ \kms, van Kempen et al.~2009). We are therefore confident that our ALMA data can be used to reliably study the  emission at
 outflow velocities greater than about 1.5~\kms.

\subsubsection{Channel Maps}
\label{chanmapsec}

 \begin{figure*}
\epsscale{1.15}
%\plotone{f1.eps}
\plotone{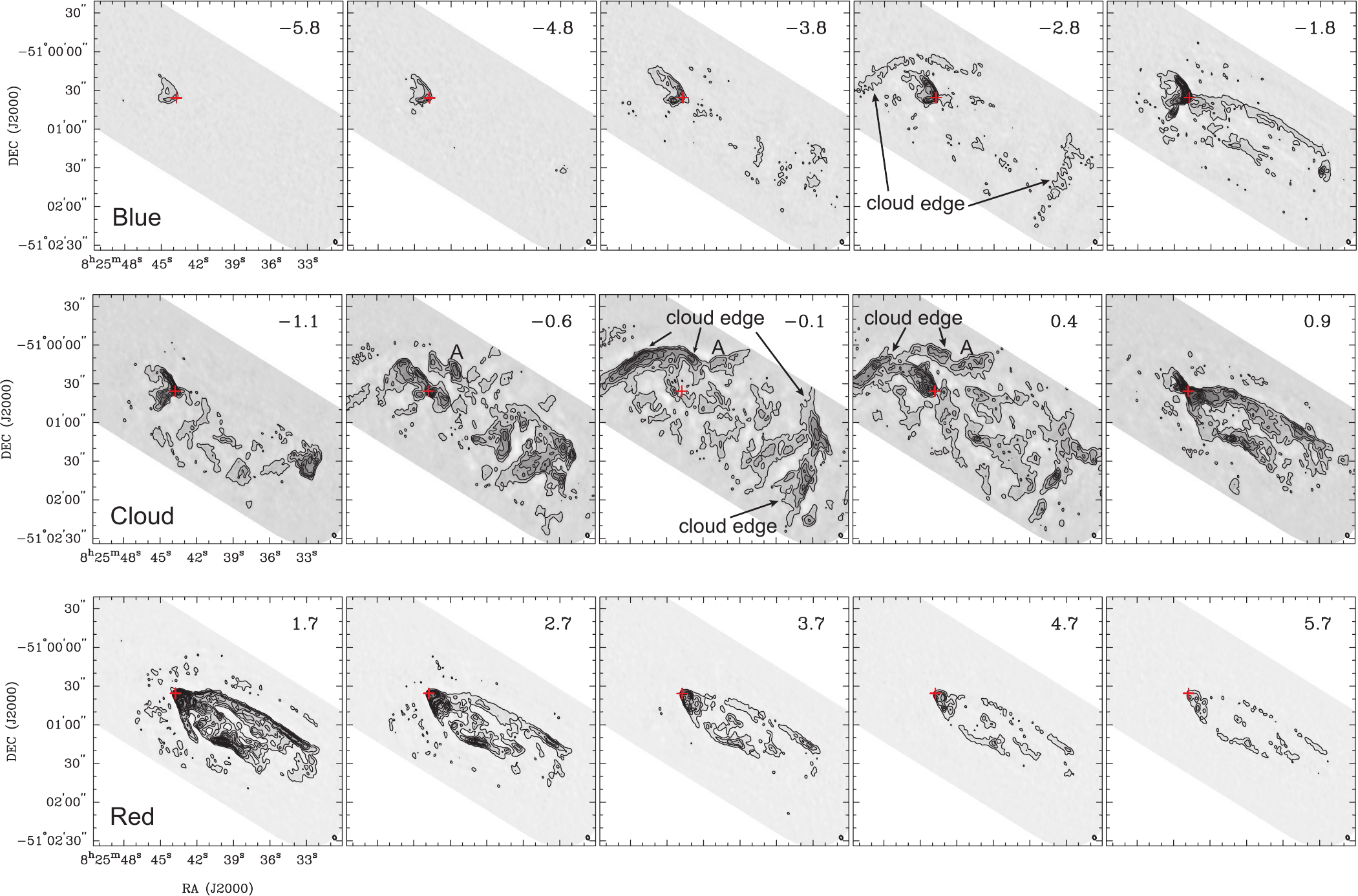}
%\vspace{-3in}
\caption{Channel maps of the CO(1--0) emission. Top (bottom) row shows blueshifted (redshifted) emission, where each panel shows the emission average over 1~\kms. The lowest contour and subsequent contour steps are 0.075  and 0.15 Jy beam$^{-1}$, respectively. The middle row shows velocities near the cloud velocity, and each panel  shows the emission average over 0.5~\kms. The lowest contour and subsequent contour steps are 0.15  and 0.3 Jy beam$^{-1}$, respectively. The central outflow velocity is given in the upper right corner of each panel in units of \kms.  The synthesized beam is shown on the lower right corner of each panel. The red cross shows the position of HH 47 IRS. 
\label{chanmapfig}}
\end{figure*}

Figure~\ref{chanmapfig} shows  the emission over the outflow velocity range between approximately $-6$ and $6$ \kms, and reveals how
the CO emission structure of the region depends on velocity. 
 We define outflow velocity  (\vout) as the LSR velocity of the emission minus the cloud LSR velocity. 
 Panels in the upper row (blueshifted velocities) and lower row (redshifted velocities) present the emission averaged over a 1~\kms \/ range centered on the velocity indicated in the upper right corner of each panel. Panels in the center row show emission averaged over 0.5~\kms \/ wide range for velocities close to that of the cloud. 
We note that low intensity outflow emission is detected at higher outflow velocities than those shown in Figure~\ref{chanmapfig}. This weak high-velocity emission is only seen when integrating the emission over a larger velocity range ($\gtrsim 5$~\kms), see Section~\ref{bluelobe} and Section~\ref{redlobe}.

In Figure~\ref{chanmapfig} the first two panels in the upper row (\vout \/ $=-5.8$ and $-4.8$~\kms) show an approximately symmetric parabolic (low-intensity) structure extending northeast of the source. At lower blue outflow velocities the emission is stronger and the northern wall of the lobe extends further from the source than the southern wall. 
At \vout \/ $=-2.8$ \kms \/ we detect two faint arcs of emission, about 40\arcsec \/ northeast and 2\arcmin \/ southwest of the protostar, which we ascribe to the limb-brightened edge of the host globule. These arcs are brighter closer to the cloud velocity (see panels at  \vout \/ $=-0.1$ and $0.4$~\kms). At \vout \/ $=-1.8$~\kms \/ the blueshifted outflow emission is very bright and it is concentrated in an inclined V-shape structure northeast of the source. It is only at very low outflow velocities that we detect the clump B0, about 20\arcsec \/ southeast of HH 47 IRS, which very likely traces the molecular outflow driven  by  the binary component that is not driving the main  HH 46/47 outflow (see Section~\ref{binary}). In this panel we also detect faint emission southwest of the outflow source,  coincident with some of the regions where we detect redshifted outflow emission.  
%In Section~\ref{bluelobe} we suggest
It is possible 
 that the walls of the southwest (mainly redshifted) outflow lobe are expanding into the surrounding cloud (similar to the RNO 91 outflow; Lee \& Ho 2005), and the filamentary blueshifted emission comes from the front walls of the cavity.  In addition, at \vout \/ $=-1.8$~\kms \/ we detect a small relatively bright clump close to the position of the HH 47C  shock 
 (and the redshifted clump R3) at about 2\arcmin \/ southwest from the source, which most likely arises from gas entrained by HH 47C.

 \begin{deluxetable*}{llccccccccc}
\tablecolumns{11}
\tabletypesize{\scriptsize}
\tablecaption{Outflows Properties  
\label{outprops}}
\tablewidth{0pt}
\tablehead{
\colhead{Lobe} &   & \multicolumn{3}{c}{Mass\tablenotemark{a} [$10^{-2}$ M$_{\sun}$]} 
&  \multicolumn{3}{c}{Momentum\tablenotemark{a, b} \/ [$10^{-2}$ M$_{\sun}$ \kms]} 
&   \multicolumn{3}{c}{Energy\tablenotemark{a, b} \/ [$10^{42}$ erg]} \\ 
&  \colhead{$T_{ex} = $} & \colhead{$15$ K} & \colhead{$50$ K} & \colhead{$100$ K} &
\colhead{ $15$ K} & \colhead{$50$ K} & \colhead{$100$ K} &
\colhead{$15$ K} & \colhead{$50$ K} & \colhead{$100$ K}
}
\startdata
%Blue & & 0.7 & 1.6 & 2.9 & 3.3 & 7.8 & 14.5 & 3.3 & 7.7 & 14.3\\
%Red & & 3.3 & 7.7 & 14.4 & 11.6 & 27.3 & 50.8 & 8.9  & 20.8 & 38.7 \\
%\enddata

Blue & & 1 & 2 & 3 & 3 / 6 & 8 / 16  & 15 / 30  & 3 / 12 & 8 / 32 & 14 / 56 \\
Red & & 3 & 8 & 14 & 12 / 24 & 27 / 54 & 51 / 102  & 9 / 36  & 21 / 84  & 39 / 156  \\
\enddata

\tablenotetext{a}{Estimates are obtained assuming emission is optically thin. 
Values should be treated as lower limits.}
\tablenotetext{b}{Values before the slash are not corrected for the outflow inclination 
and those after the slash are corrected assuming an inclination of the outflow axis, 
with respect to the plane of the sky, of $30\arcdeg$.}

\end{deluxetable*}

As shown in Figure~\ref{avespec}, close to the cloud velocity our interferometer observations do not fully recover the emission from the cloud, as the  
observations are not sensitive to large scale  ($\gtrsim 20\arcsec$) structures. 
This is the reason why the channel maps from \vout  \/ $=-1.1$ to 
$0.9$~\kms \/ appear so clumpy and the emission at these velocities cannot be used to obtain reliable estimates of the outflow or cloud mass nor to map the overall distribution of the cloud molecular gas. There are, however,  a few small-scale features that are discernible among the clumpy mess. Even 
at \vout  \/ $=-1.1$~\kms \/  we are able to identify the V-like structure just north-east of HH 47 IRS due to the blue outflow lobe, and a clump of gas close to the position of HH 47C.  At the velocity of the cloud 
(i.e., the \vout  \/ $=-0.1$ and $0.4$~\kms \/ panels) we detect emission from the edge of the cloud,  which is coincident with the diffuse 8~\micron \/ emission observed in Spitzer Space Telescope images of the region that borders the edge of the host globule 
\citep{Noriega-Crespo+04, Velusamy+07}. This structure  seems to trace the outer parts of the cloud that
are being heated by the UV radiation from the  stars that produce the Gum Nebula. 
Acceleration of the gas at the globule's edge caused by a UV-radiation-induced photoablation 
flow could explain  the fact that we detect the globule's edge at two different velocities separated by $\sim 3$~\kms \/ (i.e., at \vout  \/ $\sim 0$ and  \vout  \/ $\sim -3$~\kms). Certainly, a more detailed study of the cloud's kinematics is needed to understand the origin of the velocity structure at the globule's edge. 

The \vout  \/ $=0.4$ \kms \/ panel also exhibits a bright structure that extends northeast of the source and coincides with the HH 46 optical nebula (see Section~\ref{bluelobe}). This vey low velocity emission could be due to the gas that is entrained by the expanding back wall of the  northeast (mainly blueshifted) lobe. At \vout  \/ $=0.9$~\kms \/ the overall structure of the CO emission is similar to the integrated intensity image of the redshifted lobe in Figure~\ref{outflowmap}, yet there are some notable differences.  These include a bright narrow structure that extends  northeast of the source (coincident with the blue lobe), and the fact that the wide angle morphology of the southwest lobe is more prominent here than in the integrated intensity map (which we argue in Section~\ref{redlobe} is evidence for the existence of a wide-angle wind). 

Similar to the blue lobe, the redshifted outflow lobe is brighter at lower outflow velocities. The \vout = 1.7 \kms \/ panel shows the strongest emission, where the  highest intensity is seen close to the source and on the southern and northern walls of the outflow lobe.  In this panel the CO emission shows a wide-angle morphology close to the source, and the emission associated with the northern outflow wall is somewhat  extended (as opposed to a simple narrow structure as seen at higher outflow velocities).  The three redshifted clumps (R1, R2 and R3) are seen most clearly at \vout  \/ $= 2.7$ and $3.7$~\kms, but they are also detected  out to \vout \/ $\sim 5$~\kms. The channel maps also show that from 
\vout  \/ $=1.7$~\kms \/ to \vout  \/ $=5.7$~\kms  \/ the redshifted lobe becomes slightly narrower as the velocity increases.

\subsubsection{Outflow Mass, Momentum and Energy}
\label{outmass} 
To obtain a reliable estimate of the outflow mass we would need to know the opacity of the $^{12}$CO(1--0) line, as in outflows the opacity is expected to vary with velocity \citep{Bally+99, Yu+99, AG01b}. One way to estimate the opacity of the $^{12}$CO(1--0) line  
is using the ratio of $^{12}$CO(1--0) to $^{13}$CO(1--0). Unfortunately, there are no existing observations of the $^{13}$CO(1--0)  emission with an angular resolution (and sensitivity) similar to our 
$^{12}$CO ALMA map that would allow us to properly correct for the opacity. We thus  estimate the mass assuming the line is optically thin, and warn that our masses are very likely underestimated, by a factor of few or even as much as an order of magnitude \citep[e.g.,][]{CB90, AG01b, Offner+11}.  We follow \citet{Bourke+97} to estimate the outflow properties from the CO(1--0) data, assuming an abundance ratio of $[^{12}$CO$]/[$H$_2] = 10^{-4}$, as in \citet{Olberg+92}. 

An estimate of the CO outflow excitation temperature  is needed to calculate the outflow mass. However, there is a very wide range of temperature values given in the literature for the HH 46/47 molecular outflow.  \citet{CM91} use the ratio of brightness temperature between the CO$(3-2)$ and CO$(2-1)$ at outflow velocities to estimate an outflow excitation temperature ($T_{ex}$) of $8.5 \pm 1$~K 
(but use a value of 10 K in their outflow mass calculations). On the other hand,  \citet{Olberg+92} estimate an excitation temperature, at outflow velocities, of 15 K using the intensity ratio of CO(1--0) to CO(2--1). 
\citet{vanKempen+09}  use their multi-line observations to conduct a thorough and detailed study of the 
physical conditions of the gas surrounding HH 47 IRS, and estimate larger values for the 
outflow excitation temperature 
(between 50 and 150 K, but most likely about 100 K). 
Without ALMA observations of higher CO transition lines we cannot make our own detailed map of $T_{ex}$, 
with an angular resolution comparable to our CO data, along the entire extent of the outflow.
Therefore, we calculate the outflow mass, and other outflow properties shown in Table~\ref{outprops}, using different values of excitation temperature. It is clear that uncertainties in $T_{ex}$ result in highly  
inaccurate estimates of the outflow properties. 
We note that if we consider the same excitation temperature and the
same velocity ranges used by \citet{Olberg+92} for the two different outflow lobes 
(i.e., $T_{ex} = 15$~K, $v_{out} = 
-5.0$ to $-1.5$~\kms \/ for the blue lobe  and $v_{out} = 1.5$ to 7.5~\kms \/ for the red lobe), 
we obtain a total outflow mass of $3.6 \times 10^{-2}$~M$_{\sun}$, similar to the value of 0.03~M$_{\sun}$ obtained by Olberg et al. This is consistent with Figure~\ref{avespec}, which shows that our interferometric observations recover most (if not all) of the emission at outflow velocities greater than 1.5~\kms.

Our observations are much more sensitive than any previous study of the HH~46/47 molecular outflow, which allows us  to detect outflow emission over a much wider velocity range. 
 In this work we measure the outflow mass ($M_{out}$), momentum $[P_{out} =\Sigma ~ M_{out}(v_{out}) ~ v_{out}]$, and energy [$E_{out} = 0.5 ~ \Sigma ~ M_{out}(v_{out}) ~ v_{out}^2$]
 over the velocity range where we detect outflow emission with intensity greater than $3\sigma$, that is $-30 < v_{out} < -1.6$ \kms \/ for the blue lobe and $1.4 < v_{out} <  40$~\kms  \/ for the red lobe 
 (see Table~\ref{outprops}).
  In both lobes the very high velocity outflow  ($|v_{out}| > 20$~\kms) is constrained to within about 15\arcsec \/ of the source.
 Detection of low-intensity CO(1--0) outflow emission at very high velocities resulted in our estimates of the  
 %(by at least a factor of 4)  
 outflow kinetic energy 
 %(and to a lesser extent the
and  momentum to be significantly higher than those of \citet{Olberg+92}. 
 This implies that other similar molecular outflows may be much more energetic than what previous (low-sensitivity) observations indicated and outflows may have the potential to have more 
 impact on their surrounding cloud than previously thought. 

We can compare the properties of the jet with those of the molecular outflow in order to investigate
whether the jet injects enough momentum into the cloud and drive the molecular outflow.  
Assuming $T_{ex} = 50$~K, the momentum of the red lobe is $\sim 0.5$~M$_{\sun}$ \kms \/ 
(see Table~\ref{outprops}). The mass loss rate of the HH 47 jet (inward of HH 47A) 
was estimated by \citet{Hartigan+94} to be about $4 \times 10^{-7}$~M$_{\sun}$~yr$^{-1}$.  
Using a jet velocity  of 300~\kms \/ (see above), the momentum rate in the jet is then estimated
to be  $1.2 \times 10^{-4}$~M$_{\sun}$~\kms \/ yr$^{-1}$. 
We adopt the age of the wind to be 9000 yr, the 
dynamic age of the parsec-scale HH flow associated with HH 46/47, discovered by \citet{Stanke+99}.  
Assuming the red lobe has a similar momentum rate as the blue lobe and that the rate 
has been approximately constant over the lifetime of
the flow, we estimate that the jet has been able to inject a total of about 1 M$_{\sun}$~\kms \/ into the
surrounding medium, which could be enough to drive the molecular outflow
(if the assumptions above are correct).
However, it should be kept in mind that our estimate of the molecular outflow momentum 
%in Table~\ref{outprops} 
is probably underestimated by a factor of a few to possibly an order of magnitude
(since we assumed the emission is optically thin). Moreover, if we instead use 
 the dynamic age of the jet within the globule (using the size of the jet comparable to the molecular outflow red lobe), that is $\sim 1000$ yr,
the estimated momentum injected by the jet into the cloud would be approximately 0.1  M$_{\sun}$~\kms. 
It thus seem  
that there is material ejected by the protostar which is not detected in the optical or IR 
--and therefore not included in the estimate of the jet mass loss rate quoted above-- 
that contributes to the momentum injected by the protostellar wind into the cloud that drives the molecular outflow.

\begin{figure*}
\epsscale{1.1}
%\plotone{f1.eps}
\plotone{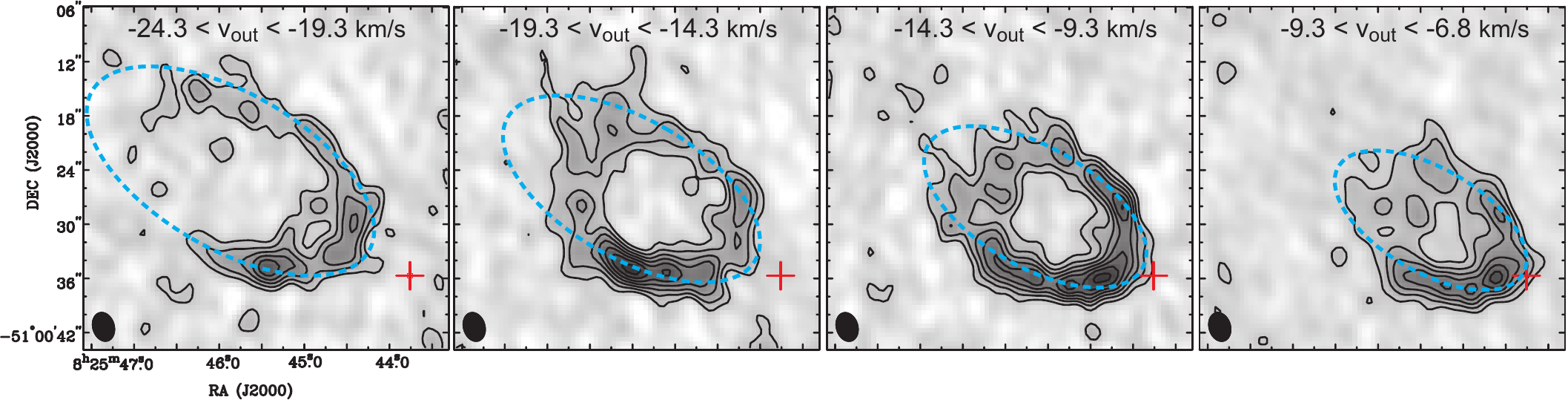}
%\vspace{-3in}
\caption{Velocity range-integrated intensity maps of the CO blueshifted  emission. The velocity interval of integration is given on the top of each panel. In all panels the lowest contour  and  contour steps are 0.07~Jy beam$^{-1}$~\kms. 
 The dash blue ellipse in each panel shows the expected shape of the molecular outflow shell driven by a  wide-angle wind using the model described in the text (see Section~\ref{bluelobe}). The model shell is
calculated using the outflow velocity at the center of the velocity interval of integration, and the 
same parameters used to model the $p-v$ diagram in Figure~\ref{pvblue}.
 The synthesized beam is shown on the lower left corner of each panel. 
The red cross shows the position of HH 47 IRS.  
\label{highblue}
}
\end{figure*}

\section{Discussion}
\label{dis}

\subsection{Blue lobe: evidence for wide-angle wind}
\label{bluelobe}

As mentioned above, we detect  blueshifted outflow emission out to larger velocities than what is presented in Figure~\ref{chanmapfig}. The intensity of the high-velocity emission is very low and it is necessary to sum over velocity ranges  wider than 1 \kms \/ in order to detect significant emission.  
In Figure~\ref{highblue} we show four velocity maps, integrated over different velocity ranges, that show the morphology of the high-velocity blue lobe gas. 
From this figure, it is clear that at higher outflow velocities the emission extends farther from the source, compared to lower velocities. In addition,  
in panels {\it b} through {\it d} the emission shows an elongated ring or shell-like structure. 
%This implies that the protostellar wind not only entrains ambient gas along
 %the  parabolic walls visible at most blue shifted velocities, but also at angles closer to the outflow axis. 
 This general behavior, in both morphology and velocity distribution, is consistent with the expected properties of a molecular outflow entrained by a wide-angle wind \citep{Shu+91, Lee+00, AG02}. 

\begin{figure}[b]
%\epsscale{1.1}
%\plotone{f1.eps}
\plotone{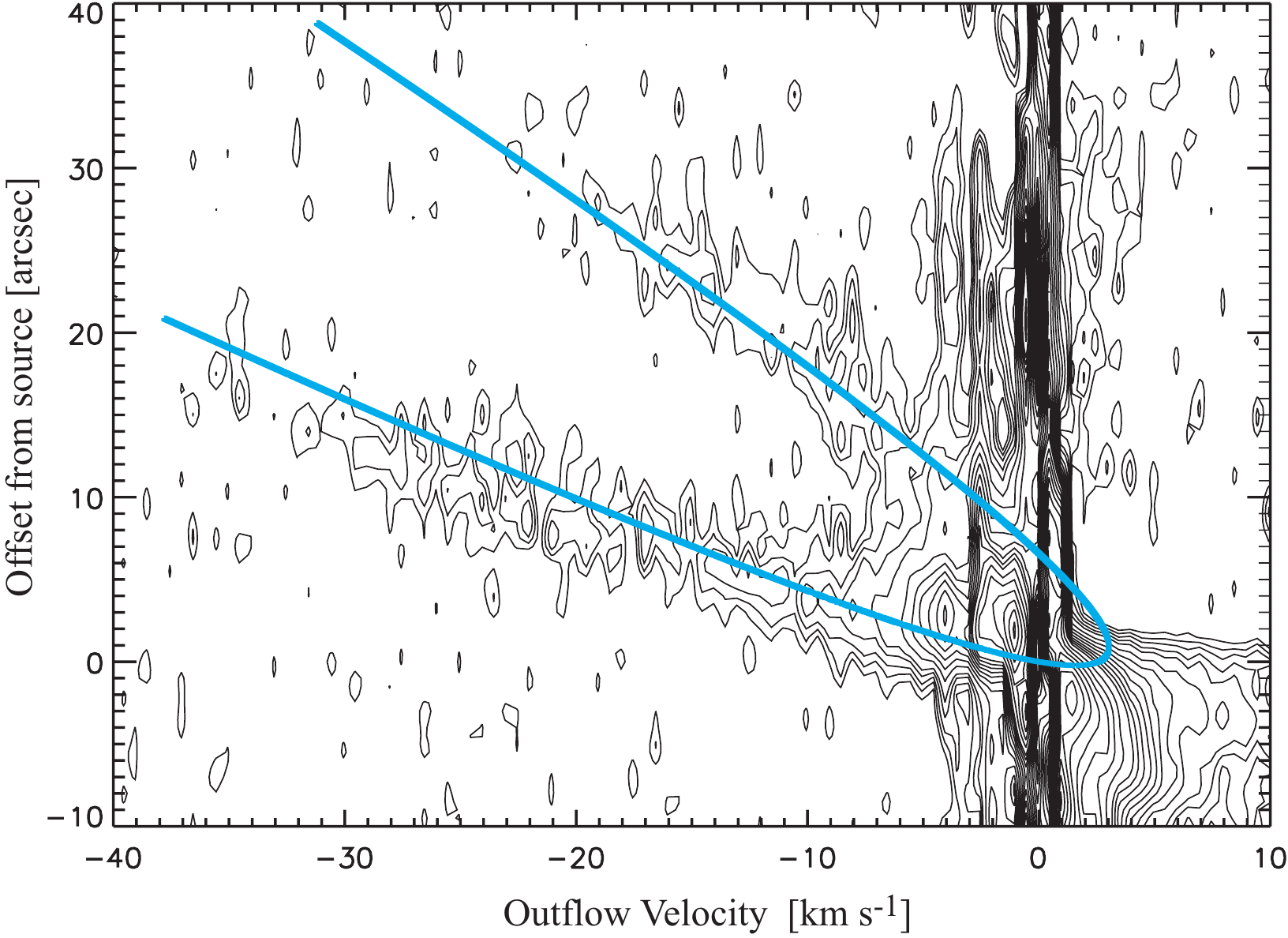}
%\vspace{-3in}
\caption{Position-velocity diagram along the axis of the blue lobe. The figure was constructed using a map with velocity resolution of 0.5~\kms, and summing the emission over a 7.5\arcsec  \/ (about 3 beams) wide
 cut, with a position angle of $60\arcdeg$.
 Contours have value of 0.35, 0.66, 1, 1.5, 2, 3 Jy beam$^{-1}$, and subsequent steps of 1 until 10 Jy beam$^{-1}$, and thereon in steps of 2 Jy beam$^{-1}$. 
The light blue parabola shows the model fit to the data (see Section~\ref{bluelobe}).     
\label{pvblue}}
\end{figure}

In order to further investigate the underlying protostellar  wide-angle wind and entrainment mechanism in the blue lobe, we constructed a position-velocity ($p-v$) diagram along the axis of the outflow, shown in Figure~\ref{pvblue}.  
This  figure %Figure~\ref{pvblue} 
shows an inclined parabolic structure, as expected  for an outflow entrained by a wide-angle wind with a non-zero inclination of the outflow axis with respect to the plane of the sky \citep{Lee+00, Lee+01}. The relatively high angular resolution and sensitivity of our ALMA data results in one of the most  (if not the most) clear case where such a parabolic structure is detected in the $p-v$ diagram of a molecular outflow lobe. 
%The structure in the blue lobe extends to outflow velocities of up to $\sim -30$~\kms.

We follow the simple analytical model by \citet{Lee+00} to describe the blue lobe of the HH 46/47 molecular outflow. In this model, 
which is based on the study by \citet{LS96}, the molecular outflow 
is made up of the gas that is swept-up by a wide-angle wind that propagates into a flattened (toroid-like) dense core. The resulting molecular outflow can then be modeled by a radially expanding parabolic shell with a velocity structure in which velocity increases with distance (i.e., a Hubble law velocity structure). 
A diagram of the model is shown in Figure 21 of \citet{Lee+00}, and in cylindrical coordinates (with 
the $z$-axis along the symmetry axis of the wind and the $R$-axis perpendicular to it) 
the morphology of the shell can be represented by the equation $z = CR^2$, while the velocity components of the shell along the $z$ and $R$ axes are represented by $v_z = v_o z$, $v_R = v_o R$. In these equations $C$ and $v_o$ are free parameters with units of arcsec$^{-1}$ and \kms~arcsec$^{-1}$,  which are constrained by the shape of the molecular outflow shell and the shape of the  $p-v$ diagram, respectively. 
%In this model the observed velocity as a function of the projected distance from the source (i.e., the apex of the parabola) is given by:
%\begin{equation}
%X=\sqrt{x^2+5}
%\end{equation}
%where $i$ is the angle between the outflow axis and the plane of the sky.

A map of the high-velocity blue lobe parabolic structure was produced by integrating the emission for outflow velocities between $-26$ and $-4$~\kms, in order to avoid ``contaminating'' emission from the cloud and the binary component outflow (see  Figure~\ref{chanmapfig}). We rotated the map by $-30\arcdeg$
 and fit a parabola ($z = CR^2$) to the structure. The fit gives a value for $C$ of $0.3\pm 0.05$, where the errors indicate the range of values that provide a reasonable fit to the map.  We then fit the $p-v$ diagram with the wide-angle
wind model of \citet{Lee+00} described above. We constrain the inclination angle with respect to the plane of the sky ($i$) to $30 \pm 10\arcdeg$, close to the values 
derived by \citet{EM94} and \citet{Hartigan+05}, assuming a distance to the parent globule of 450 pc. We find that a value of $v_o = 2.3 \pm 0.2$~\kms \/ arcsec$^{-1}$ and an value of 
$i = 29\arcdeg \pm 1\arcdeg$ result in a reasonable fit to the $p-v$ cut.  The errors indicate the range of values for which reasonable fits are attained, and choosing significantly different values for these two parameters would result in a parabolic curve with a different width or inclination compared to the $p-v$ diagram obtained from our data.

The wide-angle wind model can also be used to predict the shape of the molecular outflow shell at different velocities \citep{Lee+00, Hirano+10}.  We  used the parameters derived above, and plot the predicted shape at different outflow velocities in Figure~\ref{highblue}. For each panel %in Figure~\ref{highblue}
 the dash ellipse shows the expected shape of the molecular outflow shell, according to the wide-angle wind model,  at the outflow velocity at the center of the velocity interval of integration (shown at the top of each panel).  The model provides a reasonable match to the 
velocity range-integrated intensity maps, as the observed emission structure follows the general trend expected from the model. That is, at lower outflow velocities the outflow shell is more compact and its southwestern end coincides with the position of the source, while at higher velocities the shell is more extended and lies further away from the source.  

\begin{figure}[b]
\epsscale{1.15}
\plotone{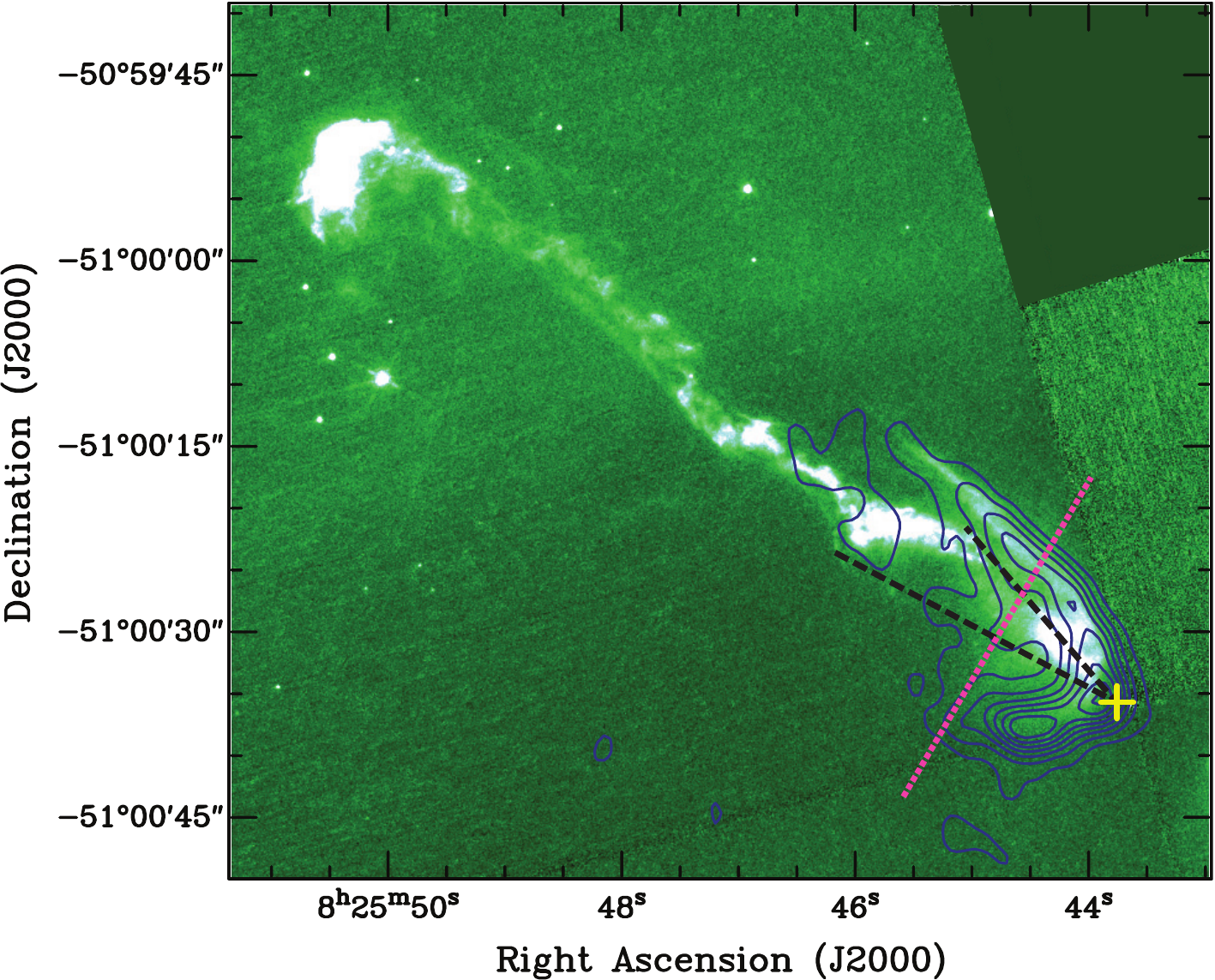}
 \caption{Comparison of blue CO lobe and optical image of HH 46/47 jet.
 The HST image was taken with the [SII] (F673N) filter in 2008 by \citet{Hartigan+11}. 
 Contours show the emission integrated over the velocity range from -6.8 to -3.2~\kms. 
  This velocity range was chosen to avoid contamination from cloud emission and B0.
%   (see~Section~\ref{chanmapsec} and Figure~\ref{chanmapfig}). 
The lowest contour and subsequent contour steps are 0.15  and 0.2 Jy beam$^{-1}$~\kms, respectively. 
  The yellow cross shows the position of HH 47 IRS.
 The two black dash lines connect the position of the source to outermost positions of the wiggling jet, with respect to the outflow axis. 
  The dotted pink line perpendicular to the outflow cavity shows the position of the $p-v$ cut shown in Figure~\ref{pvcavity}{\it a}.
\label{HSTblue}}
\end{figure}

\begin{figure*}
\plotone{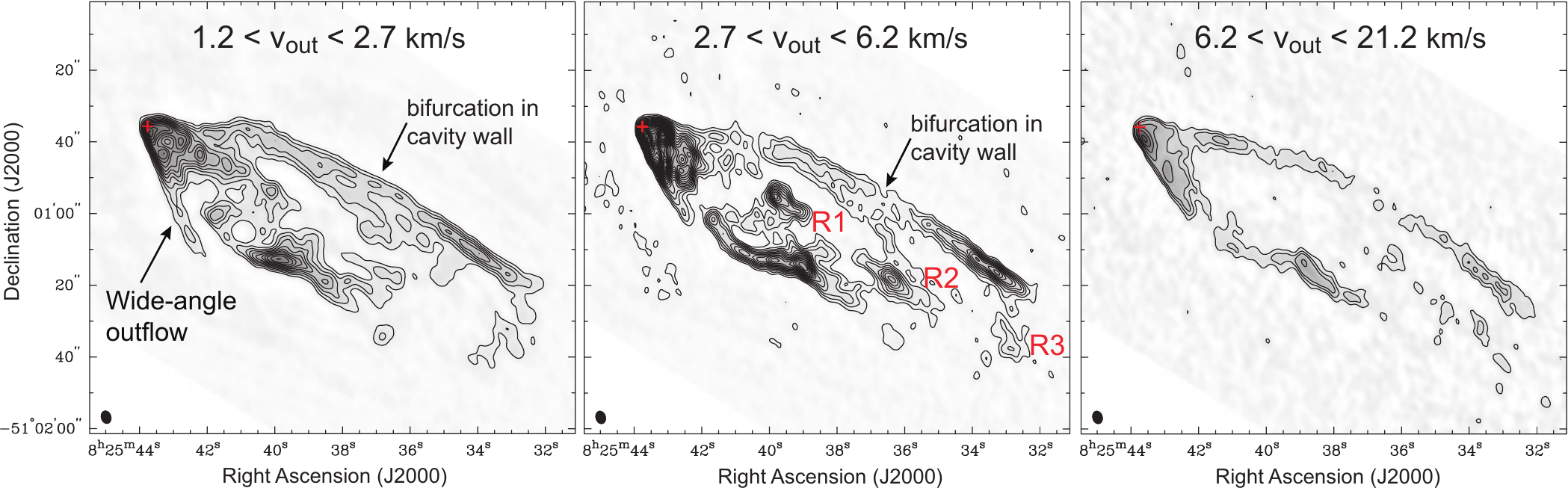}
 \caption{Velocity range-integrated intensity maps of the CO redshfited  emission. The velocity interval of integration is given at the top of each panel. In the left, middle and right panels the lowest contour and contour 
 steps are 0.2, 0.105, and 0.2~Jy beam$^{-1}$~\kms, respectively. 
 The synthesized beam is shown on the lower left corner of each panel. The red cross shows the position of HH 47 IRS. 
\label{redlobefig}}
\end{figure*}

This simple analytical model is probably not the only  model that may reproduce the observed morphology. For example, in a  jet (or collimated wind) with  varying ejection velocity, such as HH 46/47 \citep[e.g.,][]{Raga+90},  fast-moving ejecta can collide with previously ejected (slower-moving) material. As a result of such 
collisions  the wind material may expand and push ambient material in a direction perpendicular to the jet axis, 
possibly creating a wide cavity with low-collimation \citep[e.g.,][]{Suttner+97, Cabrit+97}.
However, the fact that we can reasonably fit   
 the velocity range-integrated intensity maps (Figure~\ref{highblue}) 
 and the $p-v$ diagram (Figure~\ref{pvblue})
with the wide-angle wind model described  above
suggests that the observed blue lobe could very likely have been formed by the entrainment of a wide-angle wind and the ambient cloud. 
Our data, however, cannot distinguish between possible wind-launching mechanisms that can produce wide-angle winds 
(e.g., X-wind, Shang et al.~2006 or disk winds, Pudritz et al. 2006). We would need higher angular resolution data to probe down to less than $\sim 50$ AU from the source
in order to possibly  discriminate between different launching mechanisms \citep[e.g.,][]{Ferreira+06}. 

It might seem conflicting that  the blue outflow lobe appears to be entrained by a wide-angle wind, when
 it is coincident with the base of a well-studied jet (see Figure~\ref{HSTblue}). 
 However, it is possible for both a collimated wind component (i.e., a jet) to co-exist with a wide-angle wind 
\citep[see, e.g., models by][]{Shang+07, Pudritz+07, Fendt09, Tomida+13}.
In fact, this is not the first source to exhibit such characteristics. Three examples where a ``dual-component'' wind have been invoked to 
explain the mm and optical/IR outflow observations are  HH 111 
(Nagar et al.~1997, Lee et al.~2000, however see Lefloch et al.~2007 for an alternative explanation), 
  HH 315 \citep{AG02}, and B5-IRS 1 \citep{Yu+99}.
In the first two cases the protostellar source is close to the cloud edge, the blue optical jet (or HH flow) is clearly seen to reside outside the dense parts of the cloud, and the morphology  of the molecular outflow's blue lobe is consistent with it being formed by a wide-angle wind (similar to HH 46/47). This is possible if the underlying protostellar wind has   both a collimated (jet) and a wide-angle component and the gas in the blue lobe is mostly entrained by the wide-angle component.  The reason that the wide-angle wind dominates the gas entrainment in these sources (including HH 46/47) is possibly due to the fact that the jets lie in a region of very low density molecular gas, on the outskirts of the cloud. Numerical simulations of  jet shock-driven outflows show that most of the gas entrainment takes place at (or near) the head of the bow-shock \citep[e.g.,][]{Smith+97, Lee+01}, while in a radially expanding wide-angle wind entrainment mostly takes place in a wide-angle shell originating at the source. Hence, in a dual component wind where the jet bow shocks mostly reside outside the cloud, the circumstellar molecular gas will mostly be accelerated by the wide-angle wind component close to the source, where there is enough molecular material for the wind to entrain and form the observed molecular outflow. 

Further evidence that the blue lobe of the HH 46/47 molecular outflow is 
 entrained by a dual-component wind is
observed by comparing the morphology of the  blueshifted gas and that of the optical jet. In Figure~\ref{HSTblue} we show contours of the low-velocity blueshifted lobe plotted over the HST image of the HH 47/46 jet.
The northern wall of the CO blue lobe follows the northern edge of the 
optical nebula, which implies that the bright parabolic structure observed in CO traces the walls of the outflow cavity. 
The jet's wiggling structure suggests that the jet axis changes with time or precesses   
\citep[i.e,][]{Reipurth+00}. In Figure~\ref{HSTblue} we show the opening angle of the  precession
cone in the plane of the sky, delimited by lines that connect  HH 47 IRS and the bright emission knots 
at the most extreme angles with respect to the protostellar source. It is clear that the opening angle of the molecular outflow lobe is substantially wider than that of the jet precession cone. It is therefore highly unlikely that 
precession of the jet alone could, by itself, produce the wide-angle
cavity traced by the CO outflow.

\subsection{Red lobe}
\label{redlobe}
Compared to the blue lobe, the red lobe of the HH 46/47 molecular outflow exhibits a more complex spatial and kinematical structure. In Figure~\ref{redlobefig} we show maps of the redshifted outflow emission integrated over different velocity ranges chosen to highlight the different important structures discernible at different  velocities. These are discussed in detail below. 

\begin{figure}[b]
\plotone{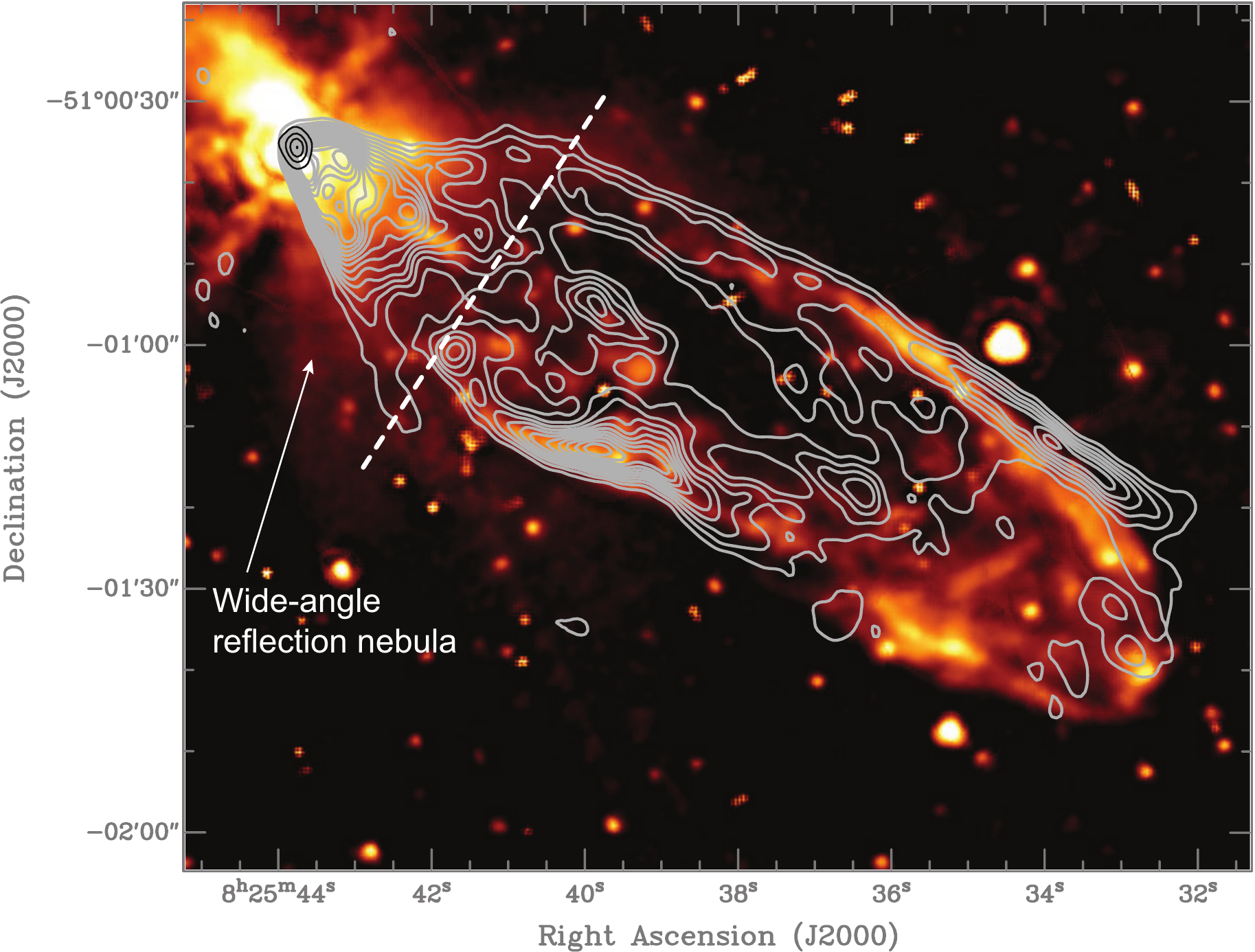}
 \caption{Comparison of red CO lobe with Spitzer IRAC 2 (4.5 \micron) image.
The Spitzer data are from \citet{Noriega-Crespo+04} and have been  reprocessed
with a deconvolution algorithm 
to reach an angular resolution of $\sim 0.6-0.8\arcsec$, with 60 iterations 
(see Noriega-Crespo \& Raga 2012 for details). White contours show the integrated intensity of the
CO redshifted lobe (the same as shown in Figure~\ref{outflowmap}). Black contours show the  3 mm continuum emission around HH 47 IRS. The lowest contour and subsequent contour steps are 2 mJy beam$^{-1}$. The
white dash line shows the position of the $p-v$ cut shown in Figure~\ref{pvcavity}{\it b}.
\label{Spitzerred}}
\end{figure}

\subsubsection{Morphological evidence for the existence of a wide-angle wind}
 Figure~\ref{redlobefig}{\it a} presents a map of the low outflow velocity emission ($1.2< v_{out} <2.7$~\kms), where the lobe shows a clear wide-angle structure, with an opening angle of $65\arcdeg$, within 40\arcsec \/ of the source. This wide-angle structure is drastically different from the morphology of the bright IR outflow cavity walls seen in the IRAC Spitzer observations \citep{Noriega-Crespo+04}, which is similar to the structure of the CO outflow red lobe at distances from the source further than 40\arcsec \/ (see Figure~\ref{Spitzerred}).  The loop-like structure seen in the IRAC images has been adequately modeled as the cavity walls of a jet-driven outflow by \citet{Raga+04}. This model, however,  does not reproduce the wide-angle structure seen in low-velocity CO emission  nor the wide-angle IR nebula seen close to the source in re-processed IRAC images of HH 46/47 (see below). It seems, therefore, that the wide-angle structure is not  produced by the collimated (jet) component responsible for the bright outflow cavity walls seen both in CO and IR, extending form about 30\arcsec  \/ to 2\arcmin \/ away from the source.
Similar to the blue lobe, the red lobe shows evidence of both a collimated and a wide-angle wind component.  
 
 Recently, \citet{Velusamy+07} presented reprocessed images of the IRAC observations of HH 46/47, 
 which included 
 suppression of side lobes from bright sources and enhanced resolution (compared to the images presented by Noriega-Crespo et al.~2004). The 3.6 
 and 4.5 \micron \/ images show  diffuse and extended emission close to the source with a parabolic morphology, with an opening angle of about 110\arcdeg, which   
  \citet{Velusamy+07}  argue is due  to scattered light arising from a wide-angle outflow cavity 
  (see Figure~\ref{Spitzerred}). 
  This structure is wider than the CO outflow lobe, and we suspect we are not able to trace the full extent of the IR reflection nebula due to the high opacity of the CO(1--0) at velocities close the ambient cloud. In the few cases where a molecular outflow
 has been mapped in $^{13}$CO, it typically shows a wider opening angle (at lower outflow velocities) than the lobe traced by the $^{12}$CO emission and the optical/IR reflection nebula \citep[i.e,][]{Ohashi+97, TM97, AS06}. We expect that future observations using lines with a lower opacity (or that probe higher densities) will be able to trace the wide angle cavity outlined by the faint extended IR emission.

 \subsubsection{Evidence for episodic mass ejections}
   Among the most distinct structures in the velocity-integrated map shown in Figure~\ref{redlobefig}{\it b} are  three bright and compact features  along the outflow axis. These local-maxima of emission  are the same  clumps identified in Figure~\ref{outflowmap}  as R1, R2, and R3. They are more clearly seen in Figure~\ref{redlobefig}{\it b} because of the limited velocity range used to produce this figure compared to Figure~\ref{outflowmap}. Table~\ref{redclumps} shows the distance from the
source as well as other physical properties of each of the molecular clumps. The distance from the source to R1 (about 0.1 pc) is approximately equal to the distance 
between R1 and R2, and the distance between R2 and R3. This suggests they might be the result of  periodic ejection episodes, similar to those found in other molecular outflows (e.g., L1448, Bachiller et al.~1990;
 RNO 43, Bence et al.~1996; HH 300, Arce \& Goodman 2001b). In fact, R1 and R2 are at similar distances from HH 47 IRS as the bright bow shock structures HH 47B and HH 47A, respectively, along the optical jet, in the opposite (blue) lobe of HH 46/47. 
These two HH knots are thought to  arise from different mass ejection episodes \citep{RH91},
and the R1 and R2 may be associated with the counter-ejections (in the redshifted lobe) of the mass-loss events that caused HH 47B and HH 47A.  
Similarly, R3 is coincident with HH 47C (in the red lobe), which is thought to be associated with the ejection responsible for the HH 47D bow shock (in the blue lobe).

\begin{figure}
\epsscale{1.17}
\plotone{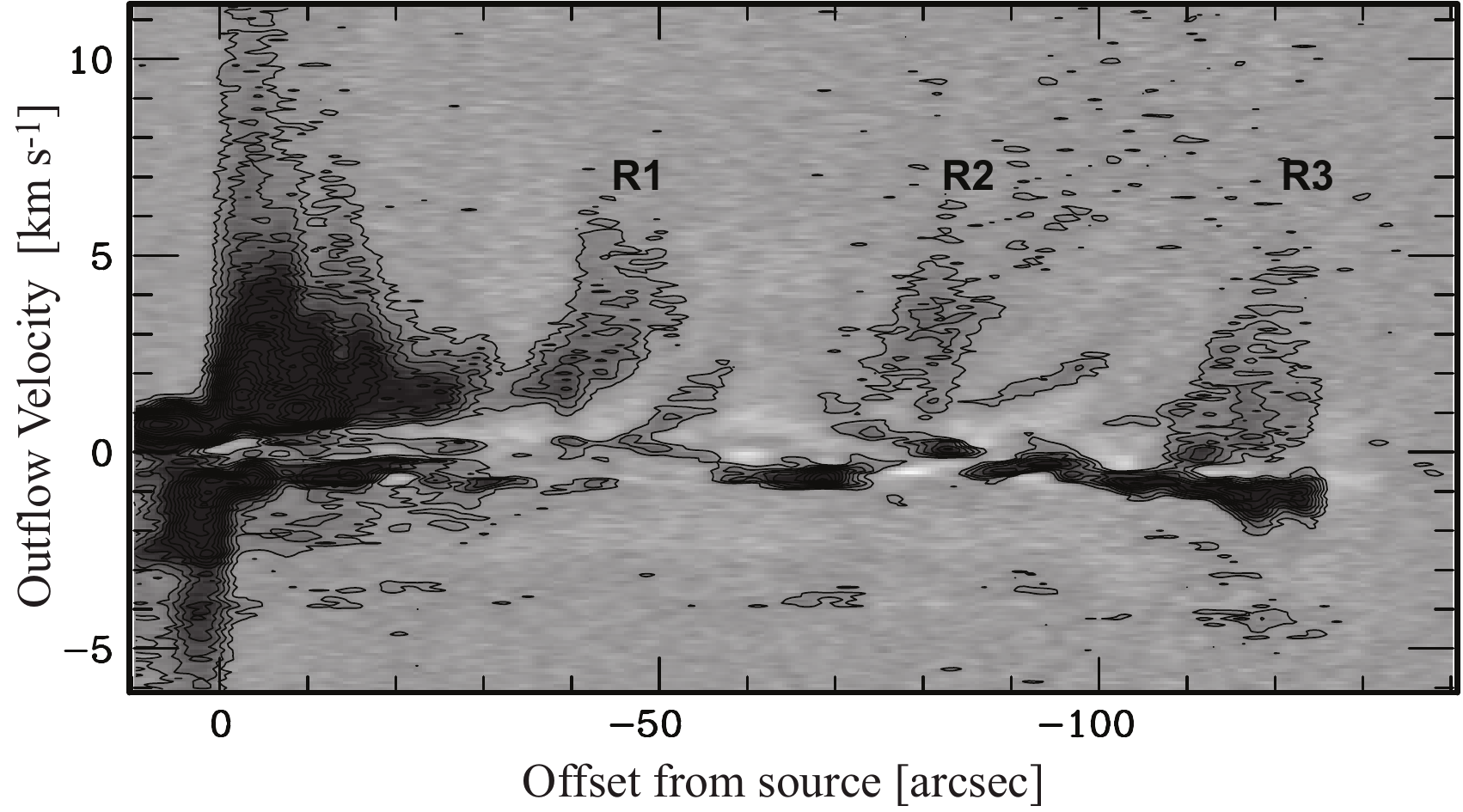}
 \caption{Position-velocity diagram along the axis of the red lobe. The figure was constructed using a map with velocity resolution of 0.08~\kms, with a position angle of $60\arcdeg$, and summing the emission over a 7.5\arcsec-wide cut. 
 The lowest contour and subsequent contour steps are 1 and 1.5 Jy beam$^{-1}$, respectively.
 The positions of the redshifted outflow clumps along the outflow axis are shown.
 \label{pvred}}
\end{figure}

 To further investigate the nature of the outflow clumps we show, in 
 Figure~\ref{pvred}, a $p-v$ diagram along the axis of the redshifted molecular outflow lobe. At distances greater than 30\arcsec \/ from the source, the $p-v$ diagram shows three regions where there is a clear increase in the outflow velocity coincident with the position of the red outflow clumps.  These features in the $p-v$ diagram, generally referred to as ``Hubble wedges'' 
\citep{AG01a},  show maximum outflow velocity increasing as a function
of distance from the source, within a localized length along the outflow axis.
% with the tip of each Hubble wedge coincident with the position of a bow shock apex or bright molecular outflow clumps.  
The general consensus is that such features are formed by 
a collimated (jet-like) episodic flow with a significantly varying mass-ejection rate, where each clump 
associated with a Hubble wedge
 is produced by the bow-shock entrainment of an individual mass-ejection  event \citep{AG01a, Lee+01}. 
 That is, each clump is mostly  made of ambient gas that has been swept up and accelerated by a 
 protostellar wind ejection episode. 
%We therefore interpret the redshifted outflow clumps in HH 46/47 as arising from different mass 
%ejection episodes from HH 47 IRS. 

In  Table~\ref{redclumps} we list 
 estimates of the timescale (i.e., age) for each mass ejection episode in HH 46/47 assuming 
 that all clumps are moving away from the source at a constant space velocity.
 We expect the ejected material will decelerate as it moves 
through the cloud and interacts with the surrounding medium, which will result  
in slower velocities for older ejected material 
\citep[e.g.,][]{CR00,GA04}. Yet, we  
can use an average flow velocity of the jet ($\bar{v}_{jet}$) for all outflow clumps 
to obtain an estimate of the ejection age for each mass ejection episode.  
Existing studies indicate that  $\bar{v}_{jet} \sim 300$~\kms  \/
for HH 46/47   
\citep{EM94,Micono+98,Hartigan+05}. Using this velocity, we estimate that a major mass ejection event in HH 46/47 takes place approximately every 300 yr, consistent with the results from studies of the optical (blueshifted) jet.

\begin{deluxetable}{lcccc}{3.5cm}
\tablecolumns{5}
\tabletypesize{\scriptsize}
\tablecaption{Properties of clumps in redshifted CO outflow lobe
\label{redclumps}}
\tablewidth{0pt}
\tablehead{    
 &  \multicolumn{2}{c}{Distance from source\tablenotemark{a}} & \colhead{Age\tablenotemark{a}} & \colhead{Mass\tablenotemark{b}}\\
\colhead{Clump} & \colhead{(arcsec)} & \colhead{(pc)} & \colhead{(yrs)} & \colhead{($10^{-3}$ M$_{\sun}$)}
}
\startdata
R1 &   42 & 0.1 & 360 &  2.1\\  %/ 11000  mass(Tex=15K) = 0.9
R2 &   81 & 0.2 & 650  & 1.8\\ %/ 20000  mass(Tex=15K) = 0.8
R3 & 118 & 0.3 & 945  & 0.7 \\ %/ 28000  mass(Tex=15K) = 0.3
\enddata

\tablenotetext{a}{Values of linear distance and age have been corrected for the inclination of the outflow axis,
assuming $i=30\arcdeg$.} 
%\tablenotetext{b}{Values before the slash are obtained using $v_{jet}=300$~\kms, 
%and those after the slash are obtained using $v_{COmax} = 10$~\kms.}
\tablenotetext{b}{Mass estimate obtained assuming optically thin emission and an
excitation temperature of 50 K. Values shown should be treated as lower limits.}  

\end{deluxetable}

The CO outflow clumps we detect are most likely made of gas that has been entrained near the current position of the clump.  Observations indicate that the outflow mass rate in HH 46/47 is of the order of $10^{-7}$ M$_{\sun}$~yr$^{-1}$ \citep{Hartigan+94, Antoniucci+08, GarciaLopez+10}. 
From the length of R2 along the outflow axis (i.e., $\sim 20\arcsec$), and assuming a constant $v_{jet} = 300$~\kms, we estimate that the episodes of high mass ejection last about 
$10^2$ yrs. With the mass outflow rate given above this would result in a total of  $ \sim 10^{-5}$ M$_{\sun}$ ejected in the wind, which is about two orders of magnitude less than the estimated gas mass of R2.    
 The mass loss rate during the episode associated with R2 would have to be about two orders 
 of magnitude  higher than the rates estimated for HH 46/47
 for the protostellar wind mass to be similar to the estimated gas mass of R2. 
Hence, it seems more likely that the vast majority of the detected molecular gas in the redshifted outflow clumps is cloud material that has been entrained by the protostellar wind rather than  material from the protostellar wind itself.

Episodes of enhanced mass outflow rate are thought to be caused by an increase in the mass accretion rate onto the protostar. In the case of HH 46/47, 
there seems to be a mechanism that is able to produce an enhancement in the accretion rate every few hundred years. This timescale is significantly shorter than the estimated orbital period of the binaries in the  HH 47 IRS system \citep[see][]{Reipurth+00}. This rules out triggering of
gravitational instabilities in the circumstellar disk of the source that drives the  HH 46/47 outflow 
by the close passage of the known binary companion as a mechanism for enhanced disk accretion. A closer
(yet unobserved) companion would have to exist for companion-disk
interactions \citep{Reipurth00} to be a feasible mechanism for triggering episodic mass ejections. Other mechanisms that could produce episodic accretion
(and hence mass ejection) events have been discussed in the literature, where disk instabilities are the main cause of enhanced accretion events \citep[e.g.,][]{VB05, Zhu+10, Machida+11}.  
In many of the  models the timescales between successive episode may be very  different depending on the assumed input parameters, and in some of these  models successive events of high mass accretion rate can occur on timescales as low as a few 100 yrs to about $10^3$~yr, consistent with the observed properties of
the HH 46/47 outflow. Future studies of outflow episodicity using a large sample may help constrain variable accretion models.

Another morphological feature detected in the redshifted molecular outflow lobe that may be caused by multiple mass ejections is
 the bifurcation of the northern cavity wall. This feature is seen at low and medium outflow velocities 
 (Figures~\ref{redlobefig}{\it a} and \ref{redlobefig}{\it b}) at about 1\arcmin \/ away from the protostar along the northern wall of the outflow lobe. Here the wall appears to divide into two filaments: a bright component that extends southwest approximately parallel to the outflow axis (PA $\sim -120\arcdeg$) and a fainter component that makes a sharp southward turn toward the position of R2 (see Figure~\ref{redlobefig}{\it a} and {\it b}). 
 %We believe that the latter feature is probably due to the same ejection episode responsible for R2. 
 Models and simulations of molecular outflows formed by episodic winds show that a thin shell or a cavity is formed around 
each internal bow shock formed by different ejection episodes \citep{Gueth+96,Lee+01}. We thus  surmise that the filamentary feature  that connects R2 and the northern wall of the red lobe traces the shell formed by the interaction of the ejection episode associated with R2 and the ambient gas.
Faint emission connecting R1 to the southern wall of the lobe may indicate that a similar structure is also driven by R1. 

%Start here XX
%9000 AU = 0.04 pc =$1.2x10^{12}$ km

 \subsubsection{High-velocity redshifted emission}
 Figure~\ref{redlobefig}{\it c} shows a velocity-integrated map over high CO outflow velocities in the red lobe ($6.2 < v_{out} < 21.2$~\kms). Here  the outflow opening angle  near the source ($\sim 50\arcdeg$) and the lobe width are  smaller than at low outflow velocities.  The overall arc-like morphology of the northern wall is consistent with the shape expected for an outflow lobe formed by (jet) bow-shock entrainment \citep[e.g.,][]{RC93, Lee+01}. 
This and the other morphological and kinematical features discussed above imply that the dominant entrainment mechanism of the  
CO outflow observed beyond 40\arcsec \/ arises from a collimated episodic wind with a series of bow shocks along the axis. On the other hand, near the source, and at  low outflow velocities the molecular outflow appears to have been formed by the entrainment from a wide-angle wind. A possible explanation is that the wide-angle component has a lower velocity (and possibly a lower density) than the collimated (jet-like and episodic) component. 
Near the source,  where the ambient density is high, both the wide-angle and collimated component will entrain the gas, but the effects of the wide-angle wind will be seen  most clearly at low outflow velocities and at relatively large angles away from the outflow axis. The faster moving (and denser) collimated component is able to travel further away from the source (and even puncture out of cloud near the position of R3), and thus the CO outflow further out from the source (and at high-velocities) will exhibit properties of it being mostly entrained by a collimated wind. 
Certainly, a successful model for the HH 46/47 molecular outflow would need to reproduce the difference in morphology between the low and high outflow velocities and at different distances from the source.

\begin{figure}[b]
\epsscale{1.2}
\plotone{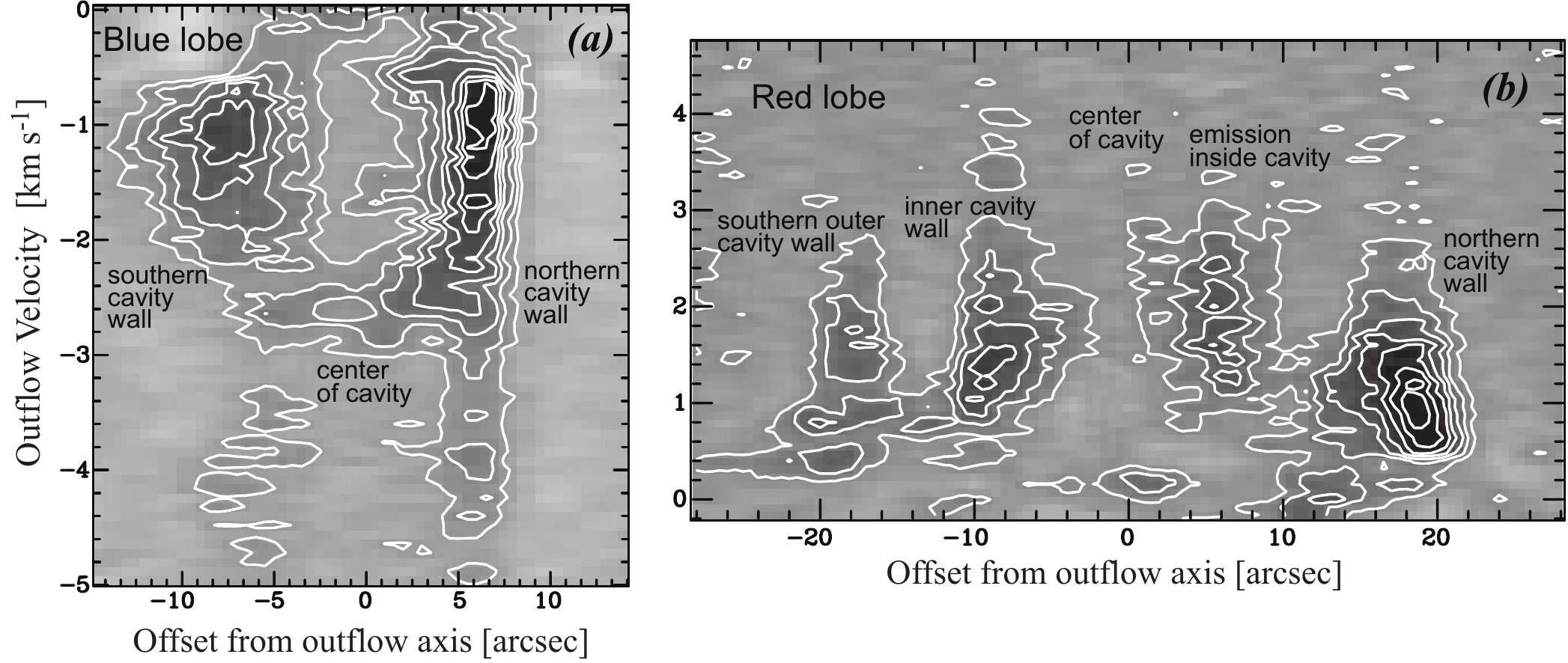}
 \caption{Position-velocity diagrams perpendicular to the outflow axis for the blue (left panel) and 
 red (right panel) lobes. In both $p-v$ diagrams the offsets are given with respect to the position
 of the outflow axis in the cut. The location of the $p-v$ cuts for the blue and red lobes are shown in Figures~\ref{HSTblue} and \ref{Spitzerred}, respectively.  
In the left panel the lowest  contour and the subsequent contour steps are 0.2 and 0.15 Jy beam$^{-1}$,
respectively. In the right panel the lowest  contour and the subsequent contour steps are 0.07 and 0.16 Jy beam$^{-1}$, respectively. 
\label{pvcavity}}
 \end{figure}
 
 \subsection{Velocity Gradients in Outflow Cavity}
The high sensitivity, high velocity resolution data allowed us to study the kinematics of the outflow cavity in detail. In Figure~\ref{pvcavity} we plot the $p-v$ 
diagrams made by cuts perpendicular to the outflow cavities of the blue and red lobes. The figure shows that  for both cavities there is a wide range of velocities present  at the position of the cavity walls, and a gradual velocity gradient with most emission near the outflow axis at higher velocities than (most of) the emission closer to the outflow walls.  In the blue lobe, at the position of the outflow cavity walls, there is detectable CO at velocities from  $v_{out} \sim -5$~\kms \/ to $v_{out} \sim -0.5$~\kms. Yet, most of the emission at these position is at velocities close to that of the ambient cloud (i.e., between \vout \/ of about  $-2$ and $-0.5$~\kms, see Figure~\ref{pvcavity}{\it a}). This in contrast to positions closer to the outflow axis, where the bulk of the emission is at slightly higher velocities, between \vout \/ of $-2$ and $-3$~\kms. A similar velocity distribution, but with much faster velocities, has been observed in the HH 46/47 optical jet with fast radial velocities  ($\sim -100$ to $-150$~\kms) in the jet core and slower velocities ($\sim -25$ to $-50$~\kms) at the edges \citep{Morse+94}.  
In the  CO outflow red lobe we see a spread in velocities (i.e., emission over a range of \vout \/ of about 2~\kms)  at the position of the cavity walls (and in the emission inside the outflow cavity), as well as an increase in the average (and minimum) velocity of the CO as the distance to the center of the cavity decreases: 
from an average velocity close to that of the ambient cloud (\vout $\sim 0.5$~\kms)
 at the outer cavity wall,  to  an average \vout $\sim 2$~\kms \/ at center of the outflow lobe  
 (see Figure~\ref{pvcavity}{\it b}). 

The wide spread in velocities at the position of the cavity walls may be explained by both geometric effects and the expected interaction between the outflow and the surrounding cloud.  Observations with limited angular resolution are expected to detect a large range in  radial velocities (i.e., the component of the gas velocity along the line-of-sight) at the limb-brightened edge of an expanding parabolic (or cylindrical) cavity, as outflow gas from the front and back side of the cavity wall should contribute to the total detected emission.  In addition, given that the cavity wall is where the outflow-cloud interaction takes place, one would expect the gas to be more turbulent here than close to the cavity center, and hence show a wider spread in velocities.   
%(where the gas motions are dominated by the directional flow of the outflow) 

It is not clear what causes the observed velocity gradient perpendicular to the outflow axis.  The decrease in velocity of about 1 to 2~\kms \/ from the center of the cavity to the cavity walls  may be due to different processes. 
One possibility is that this behavior in the CO outflow results from the entrainment of a protostellar wind that has  higher velocities towards the wind axis.
For example, a disk wind where the velocity of the wind depends on the radius at which is launched \citep{Pudritz+06} or a jet made of many  small bow-shocks where the entrained material at the head of the bow shocks (along the center of the jet) is expected to have higher velocities than the material entrained at the side wings of the bow shock 
\citep{Heathcote+96}. Another alternative is that the velocity gradient results from the momentum-conserving wind entrainment that produces the molecular outflow, where lower CO outflow velocities are expected  
in denser regions near the cavity walls  \citep[e.g.,][]{AS04}. 
An additional possibility is that this is an effect due to the geometry of an expanding parabolic, conical or cylindrical cavity inclined with respect to the plane of the sky. %, where CO emission with larger components of the velocity along the line of sight are preferentially detected closer to the center of the cavity. 
 Undoubtedly, the detected gradient is not due to outflow rotation, as it has been recently claimed for a few other sources \citep[e.g.,][]{Lee+09,Zapata+10,Choi+11,Pech+12}. The observed decrease in velocity from the outflow axis towards the cavity walls is in contrast with the expected observational signatures of outflow rotation:  a gradient across the entire lobe length with one side of the lobe showing bluer (or redder) velocities with respect to the other side of the lobe \citep[e.g.,][]{Launhardt+09}.

\subsection{Outflow from binary}
\label{binary}

\begin{figure}[t]
\epsscale{1.1}
\plotone{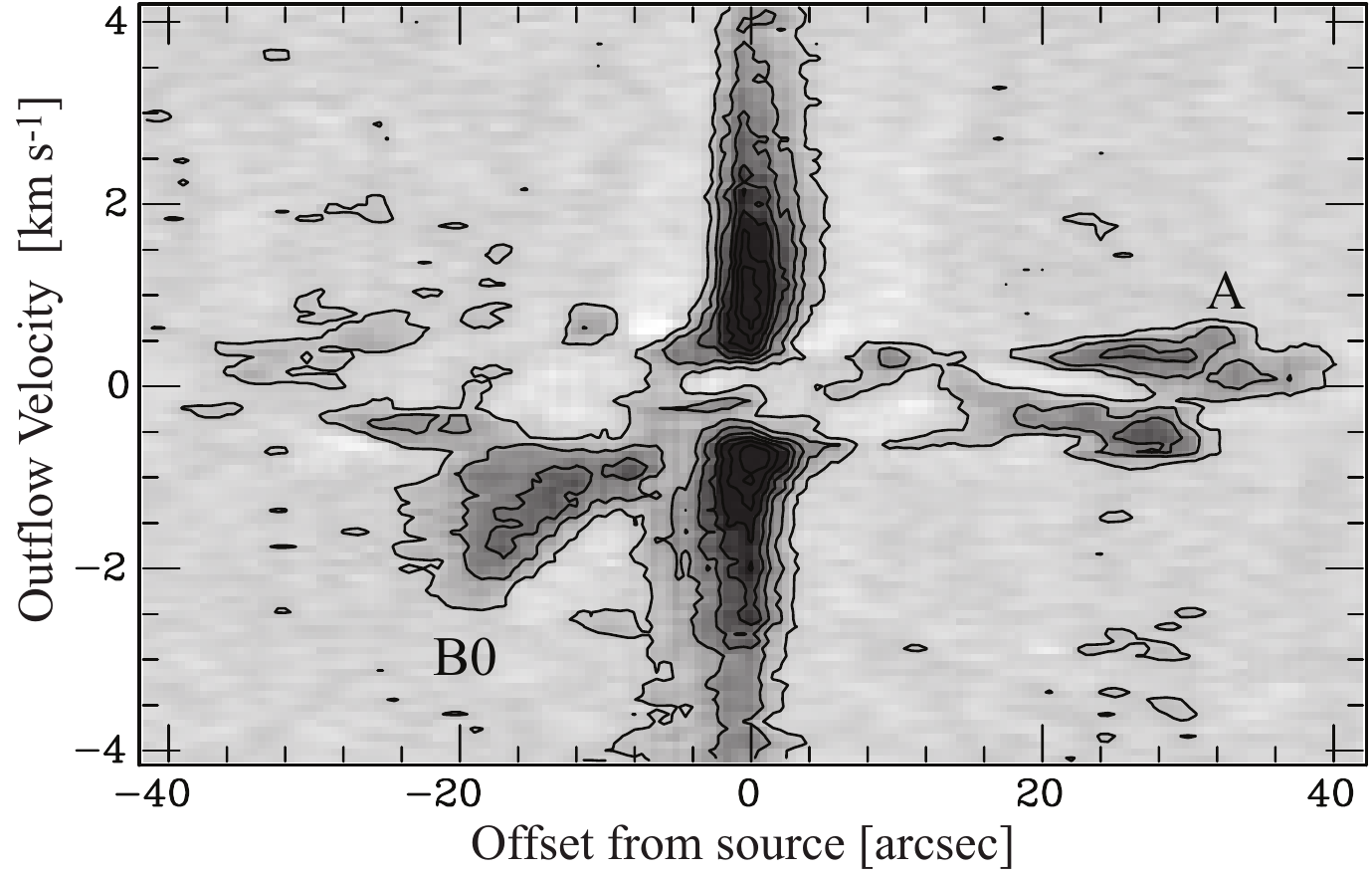}
 \caption{Position-velocity diagram along the axis of the proposed outflow powered by the
 binary companion in the HH 47 IRS system not responsible for the main HH 46/47 outflow. The location
 of the $p-v$ cut is shown in Figure~\ref{outflowmap}.
 The lowest contour and the subsequent contour steps are 0.1 and 0.3 Jy beam$^{-1}$, respectively.
 The locations of the blueshifted outflow clump B0 and clump A, along the $p-v$ cut, are shown.
\label{pvbinary}}
\end{figure}

As described in Section~\ref{chanmapsec}, the channel maps show a protuberance extending about 
20\arcsec \/ southeast of the source, at low blueshifted outflow velocities (\vout \/ $\sim -1.8$~\kms). We believe that this clump  (which we name B0) is not related to the main blueshifted outflow lobe, as it does not follow  
the general parabolic shape seen at most blue shifted velocities. 
 A $p-v$ diagram oriented along a line from HH 47 IRS to B0, at PA $\sim 132\arcdeg$ (Figure~\ref{pvbinary}), shows that the maximum velocity increases from near the source to the position of B0, similar to the Hubble-wedges of the red outflow clumps. We speculate that B0 traces a region where an underlying protostellar
  wind is entraining molecular gas. 
 
 A recent study of the HH 46/47 jet shows that the proper motion on the plane of the sky of one of the most recent ejections (within $\sim 3\arcsec$ of HH 47 IRS) is along a line with a position angle of about $40\arcdeg$ on the plane of the sky, while further 
away ($\sim 35\arcsec$) from the source  the jet motion has a PA closer to $55\arcdeg$ \citep{Hartigan+11}. Thus, it is very unlikely that the ejection responsible for B0 is part of the flow responsible for the HH 46/47 jet, and  the most likely candidate for powering the outflow that drives B0 is the other binary companion  in the HH 47 IRS binary system not responsible for the main HH 46/47 outflow.  Weaker and smaller outflows from binary components have been observed in other sources (e.g., HH 111, Cernicharo \&  Reipurth 1996; BHR 71, Parise et al.~2006; L1448C Hirano et al.~2010; IRAM 04191+1522, Lee et al.~2005, Chen et al.~2012), and we believe that HH 46/47 shows another case of drastically different outflows arising from binaries in the same system.  Current observations of this system do not have the required angular resolution to resolve which outflow is powered by each of the binary components in HH 47 IRS.

Figure~\ref{pvbinary} shows that there is a clump (which we name A) with very low red (and blue) outflow velocities at about 30\arcsec \/ northwest of HH 47 IRS on what would be the counter lobe of the outflow associated with B0. The distance from the source to A is slightly more than the distance to B0, and the position of A is coincident with the edge of the globule seen in our CO channel maps (Figure~\ref{chanmapfig}). It is not clear if clump A arises from the counter ejection of B0 or if it is just cloud emission. Further observations are needed to ascertain the nature of this clump.

%\subsection on cloud?

\section{Summary \& Conclusions}
\label{summary}

We present the first interferometric map of the molecular (CO) outflow associated with the well-known HH 46/47 flow, powered by HH 47 IRS. 
With our high velocity  ($0.08$~\kms)  and angular  ($\sim 3\arcsec$) resolution  
ALMA Cycle 0 observations we have investigated the kinematics and morphology of the molecular outflow. 
The $4.2\arcmin$ long mosaic map of the CO(1--0) emission covers the length of the HH 46/47 outflow from HH 47A, in the northeast, to HH 47C, in the southwest, and provides the data to study both the blueshifted and 
redshifted molecular outflow lobes in detail. Our main conclusions are as follows.

 The unprecedented sensitivity of our data enable us to detect outflow emission at much higher velocities than 
previous CO(1--0) studies of the source;  up to outflow velocities of about  $-30$ and 40~\kms \/ in the blue and red lobes, respectively. 
Even though the very high velocity  ($|v_{out}| > 20$~\kms) emission 
is constrained to about 15\arcsec \/ (or 7000 AU) from the source, detection of this outflow emission 
results in  significantly higher values of the outflow kinetic energy and momentum, 
compared to results from existing single dish CO(1--0) observations. 
If HH 46/47 is representative of molecular outflows from low-mass stars, our results imply that 
other similar molecular outflows may be much more energetic and carry more momentum 
than previously thought. 
The HH 46/47 flow is the first molecular outflow from a low-mass star to be mapped in the  
ALMA Cycle 0 phase and subsequent ALMA observations using more antennas will provide even
more sensitive maps. From our results we expect that future outflow studies 
with ALMA will show that outflows can have much more impact
 on their surrounding cloud than previously thought. 

The blue and red molecular outflow lobes show very distinct morphologies and kinematics. The blueshifted lobe only extends up to about 30\arcsec \/ (14000 AU) northeast from HH 47 IRS.  
The compact size of the molecular outflow lobe is due to the fact that the outflow-powering 
protostar resides at the edge of its parent globule and the protostellar wind is only able to push and 
accelerate a limited amount of molecular material before it leaves the cloud. 
The blueshifted lobe shows a very clear parabolic structure in both the integrated intensity map and in the 
position-velocity diagram along the outflow axis. The morphology and velocity distribution are consistent
with a model where the molecular outflow is formed 
by the entrainment of an underlying wide-angle protostellar wind.
 We argue that the co-existence of a wide-angle molecular outflow and an optical jet in the blue lobe of HH 46/47 is possible  if the underlying protostellar wind has both a collimated (jet) and a wide-angle component.
In the blue lobe, the wide-angle wind component dominates the gas entrainment because the jet bow shocks lie in a region with very low  ambient density while the wide-angle component interacts with  
the environment very close to the source, where there is enough ambient (molecular) gas for the
wide-angle wind to entrain and form the observed molecular outflow.

The redshifted lobe is about 2\arcmin \/ ($\sim 0.3$ pc) in length, extending out to the southwest edge of the globule. Within approximately 40\arcsec \/ (18000 AU) of the source the red lobe shows a wide-angle structure which is very different from the loop-like structure seen in IR images and 
the CO lobe structure at greater distances from the source.  
Our results also show that, beyond 40\arcsec \/ and along the axis of the red lobe, the HH 46/47 outflow has three clumps clearly detected in the integrated intensity map
and position-velocity diagram.
Their spatial distribution and velocity structure indicate that these
clumps arise from prompt entrainment, most probably produced by bow shocks,
arising from quasi-periodic episodes of increased mass ejection in HH 47 IRS. 
We argue that, similar to the blue lobe,  the
underlying protostellar wind has both a collimated and a wide-angle component  
 and that
close to the protostar the molecular outflow red lobe is dominantly driven by the wide-angle component 
while the CO outflow observed beyond 40\arcsec \/ mainly arises from the entrainment by a collimated episodic wind. 
The wide-angle wind of this, or any other, outflow may
affect the infall and the protostar's mass-assembling process, and future ALMA 
observations with higher angular resolution and using higher density tracers should concentrate 
on studying the impact of the wide-angle wind on the infalling core and circumstellar envelope. 
 
 The position of the three molecular outflow clumps along the axis of the redshifted lobe suggests that they
 are caused by the same ejection event that produced three major bow shocks in the blue lobe observed in 
 the optical. Our results confirm previous studies that indicate that  HH 47 IRS goes through episodes of  increased  mass outflow (and accretion) rates every few hundred years. Future (higher resolution) multi-epoch observations with ALMA may be able to detect the proper motion of the bow shock-driven outflow clumps in HH 46/47 and other molecular outflows. These studies could help in our understanding of outflow variability and constrain variable accretion models. 
   
Our high velocity resolution data allowed us to study the kinematics of the outflow cavity and search for 
signatures of outflow rotation. We find, in both blue and red lobes, there is outflow emission over a
wide range of velocities at the cavity walls, and the existence of a gradual gradient 
where the velocity increases towards the outflow axis. It is not clear what produces this velocity structure.
Yet, we are certain it is not consistent with outflow rotation. Future comparison of our data with numerical models will allow us to investigate the origins of this kinematic behavior in the entrained gas. 

Our CO(1--0) data detect a small clump at blueshifted velocities southeast of HH 47 IRS, which we believe arises from the interaction of a protostellar wind driven by the binary component in HH 47 IRS
that does not drive the main HH 46/47 outflow. The clump, which lies at a position angle very different from that of the 
main HH 46/47 flow, has a velocity distribution consistent with it being formed by bow shock prompt entrainment. Future observations with an order of magnitude higher angular resolution than the maps presented here will be able to confirm our results and  
resolve which outflow is powered by each of the binary components in HH 47 IRS.

\acknowledgments

HGA acknowledges  support from his NSF CAREER award AST-0845619. 
DM and GG gratefully acknowledges support from CONICYT project BASAL PFB-06.
We thank Chin-Fei Lee for his help with the wide-angle model. 
This paper makes use of the following ALMA data: 
ADS/JAO.ALMA\#2011.0.00367.S. ALMA is a partnership of ESO (representing 
its member states), NSF (USA) and NINS (Japan), together with NRC 
(Canada) and NSC and ASIAA (Taiwan), in cooperation with the Republic of 
Chile. The Joint ALMA Observatory is operated by ESO, AUI/NRAO and NAOJ.
The National Radio Astronomy Observatory is a facility of the National 
Science Foundation operated under cooperative agreement by Associated 
Universities, Inc.

{\it Facility:} \facility{ALMA}.

\end{document}